\documentclass{article}
\usepackage{graphicx, url}
\usepackage{amsmath,amsthm}
\usepackage{amssymb}
\usepackage{verbatim}
\usepackage{tikz,tikz-cd}
\usepackage{braket, url, hyperref}
\usepackage{natbib}
\usepackage{adjustbox}

\usepackage[textsize=tiny]{todonotes}

\theoremstyle{definition}
\newtheorem{define}{Definition}

\title{Is string field theory background independent?}
\author{Bhanu Narra\footnote{bhanu.narra@pmb.ox.ac.uk},~ James Read\footnote{james.read@philosophy.ox.ac.uk},~ \& Matěj Krátký\footnote{matej.kratky@etu.unige.ch}}
\date{}

\begin{document}

\maketitle

\begin{abstract}
    String field theory is supposed to stand to perturbative string theory as quantum field theory stands to single-particle quantum theory; as such, it purports to offer a substantially more general and powerful perspective on string theory than the perturbative approach. In addition, string field theory has been claimed for several decades to liberate string theory from any fixed, background spatiotemporal commitments---thereby (if true) rendering it `background independent'. But is this really so? In this article, we undertake a detailed interrogation of this claim, finding that the verdict is sensitive both to one's understanding of the notion of background independence, and also to how one understands string field theory itself. Although in the end our verdicts on the question of the background independence are therefore somewhat mixed, we hope that our study will elevate the levels of systematicity and rigour in these discussions, as well as equip philosophers of physics with a helpful introduction to string field theory and the variety of interesting conceptual questions which it raises.
\end{abstract}

\tableofcontents

\section{Introduction}
Background independence is a (supposed) theoretical virtue that, as the name suggests, aims to capture whether a given physical theory depends on or involves the choice of any fixed background structure. However, despite there being clear paradigmatic cases of both background independent and background dependent theories, stating the precise definition of background independence has proved rather challenging. One (if not \textit{the}) paradigmatic case of a background independent theory is general relativity, which can be formulated on arbitrary Lorentzian manifolds dynamically coupled to stress-energy content via the Einstein equations and which hence, intuitively, does not involve a choice of fixed background structure. On the other hand, special relativity, at least in its geometrical formulation, requires Minkowski spacetime---clearly a fixed background structure---and hence is considered background dependent.\footnote{There are subtleties here regarding different ways of formulating special relativity, but we'll elide them. For further discussion, see \citet{Pooley, ReadBI}.} Clarifying the definition of background independence is particularly urgent in light of the authority it has acquired in quantum gravity circles.\footnote{Fortunately, there is by now a lot of philosophical work that aims to fill this gap (see, e.g., \citet{ReadBI}).}

In this article, we consider the background independence of string field theory (SFT), one such candidate for a theory of quantum gravity. In the first approximation, one can think of the relationship between SFT and perturbative string theory as being analogous to the relationship between relativistic QFT and single-particle quantum mechanics.\footnote{Sometimes physicists call SFT the \textit{second-quantized} version of perturbative string theory.} While perturbative string theory describes vibrational excitations of a single relativistic string, SFT aims to describe excitations of a \textit{string field}. In assessing the background independence of SFT, we draw heavily on the physics literature. In particular, we discuss (a) the series of mathematical results by \citet{ZweibachSen, ZweibachSen2} and \citet{Sen2018} that allegedly demonstrate the background independence of SFT, (b) the `manifestly background independent' formulation of SFT due to \citet{Witten}, and (c) a contemporary extension by \citet{Ahmadain:2024hdp}. We highlight some limitations of these mathematical results, and evaluate the background independence of SFT with respect to the various definitions of background independence found in \citet{ReadBI}. In the end---as one might expect!---we find that the question of whether SFT is background independent depends both on how one understands `background independence' and on how one formulates SFT.

Here's our plan for the article. In \S\ref{Sec Worldsheet}, we provide the necessary technical background on worldsheet (perturbative) string theory; in \S\ref{Sec StringFieldT}, we do the same for (super)string field theory. In \S\ref{Sec BI}, we review arguments for SFT's background independence, and in \S\ref{Section: Invariant Structure} we present the structures which are invariant between SFTs. Doing the latter will be crucial for a full assessment of the background independence of SFT---a task we take up in \S\ref{Sec AssessingBI}.




\section{Worldsheet string theory}\label{Sec Worldsheet}

In this section, we introduce worldsheet string theory as a stepping stone to our subsequent discussion of SFT. We follow the conformal field theory (CFT) approach to worldsheet string theory (see \citet{Polchinksi}) that is best suited to understanding the formulation of covariant SFT. In \S\ref{SSec StringsinBG}, we provide the relevant background on the physics of classical strings situated in some ambient `target space' with background fields. In \S\ref{sec:worldsheet}, we review relevant details of thinking about string theory as a 2D CFT on the worldsheet. 


\subsection{Classical relativistic strings in background fields}\label{SSec StringsinBG}
The starting point for worldsheet string theory is the study of classical relativistic string propagating in Minkowski spacetime. Let $(\mathbb{R}^D,\eta)$ be a $D$-dimensional Minkowski spacetime which we call \textit{target space} and let $\Sigma_g$ be a Riemann surface of genus $g$ which we call the \textit{worldsheet}. Suppose there is an embedding $X:\Sigma_g\rightarrow \mathbb{R}^D$ which may also be specified by $D$ independent functions $X^{\mu}:\Sigma_g\rightarrow \mathbb{R}$ where $\mu$ runs from $0$ to $D$ and labels global coordinates in target space. Any such embedding describes a kinematically possible evolution of the vibrating classical string in target space. The dynamically possible embeddings are then picked out by equations of motion which follow from the Polyakov action,
\begin{equation}\label{eqn Polyakov}
    S_{\text{P}}[X^{\mu},h_{ab}] = -\frac{1}{4\pi} \int_{\Sigma_g} d^2 \sigma \ \sqrt{-h} h^{ab} \partial_a X^\mu \partial_b X^\nu \eta_{\mu\nu},
\end{equation}
where $h_{ab}$ is a Lorentzian metric on the worldsheet metric and $h :=\det \left( {h_{ab}} \right)$ ($h^{ab}$ is its inverse). The Polyakov action is highly symmetric: it is invariant under both worldsheet reparametrizations and Weyl rescalings of the worldsheet metric. More explicitly, if $f:\Sigma_g\rightarrow\Sigma_g$ is a diffeomorphism and if $\omega\in C^{\infty}(\Sigma_g)$ is a smooth function, then
\begin{align}\label{eqn invariance}
    S_{\text{P}}[f^*X^{\mu},f^*h_{ab}]=S_{\text{P}}[X^{\mu},h_{ab}] \quad \text{and} \quad S_{\text{P}}[X^{\mu},e^{2\omega}h_{ab}]=S_{\text{P}}[X^{\mu},h_{ab}]
\end{align}
express invariance under worldsheet reparametrization and Weyl rescalings, respectively.

Note that $S_{\text{P}}$ already involves a choice of background $(\mathbb{R}^D,\eta)$. However, different background choices are possible. For instance, one can ask how classical relativistic strings propagate on general Lorentzian manifolds $(M,g)$. Once again, suppose that we have an embedding $X:\Sigma_g\rightarrow M$. The natural generalization of $S_P$ is
\begin{equation}\label{eqn S_1}
    S_1[X^{\mu},h_{ab}] = -\frac{1}{4\pi} \int_{\Sigma_g} d^2 \sigma \ \sqrt{-h} h^{ab} \partial_a X^\mu \partial_b X^\nu g_{\mu\nu}(X),
\end{equation}
where $g_{\mu\nu}$ are the metric components.\footnote{There also exists a coordinate-free definition of the Polyakov action; however, its explanation involves some technicalities. Let $[\Sigma_g,M]$ be the space of smooth maps from $\Sigma_g$ to $M$ and let $\text{Met}(\Sigma_g)$ be the space of metrics on $\Sigma_g$. Associated to the embedding $X:\Sigma_g\rightarrow M$ is the tangent map $T_pX:T_p\Sigma_g\rightarrow T_{X(p)}M$ at every $p\in\Sigma_g$ and also the map of tangent bundles $TX:T\Sigma_g\rightarrow TM$. Because both $\Sigma_g$ and $M$ come equipped with Lorentzian metrics, we may define the \textit{adjoint} of $T_pX$ as the map $T_pX^*:T_{X(p)}M\rightarrow T_p\Sigma_g$ such that for any two vectors $v\in\Gamma(T_p\Sigma_g)$ and $w\in\Gamma(T_{X(p)}M)$ the relation $h_p(v,T_{X(p)}X(w))=g_{X(p)}(T_pX(v),w)$ holds. One may then define the norm $|T_pX|^2=\text{Tr}(T_pX^*\circ T_pX)$ at $p$ and the \textit{global function} on $\Sigma_g$ given by $|TX|^2(p)=|T_pX|^2$. Moreover, the metric $h$ also induces a unique volume-form $\text{vol}(h)$ on $\Sigma_g$ so that the Polyakov action may be defined as a functional $S_P:[\Sigma_g,M]\times\text{Met}(\Sigma_g)\rightarrow  \mathbb{R}$ given by $S_P[X,h]=\int_{\Sigma_g}|TX|^2\text{vol}(h)$. For more details, see \citet[Lecture 4]{Dolgachev}.}

Following the lead of classic texts such as \citet{GSW}, we also consider further generalizations of $S_{\text{P}}$ which couple the string to the remaining massless fields of the closed string spectrum: the two-form gauge field known as the Kalb--Ramond field and the scalar dilaton field.
The action terms corresponding to these couplings are respectively
\begin{align}\label{eqn S_2 and S_3}
    S_2[X^{\mu},h_{ab}] &= -\frac{1}{4\pi} \int_{\Sigma_g} d^2\sigma\ \epsilon^{ab}  \partial_a X^\mu \partial_b X^\nu B_{\mu\nu}(X), \\ \label{S_3}
    S_3[X^{\mu},h_{ab}] &= \frac{1}{4\pi} \int_{\Sigma_g} d^2\sigma\ \sqrt{-h} \Phi(X)R,
\end{align}
where $\epsilon^{ab}$ is the antisymmetric tensor density, $R$ is the worldsheet Ricci scalar, and $\Phi$ and $B_{\mu\nu}$ are some fixed background choices for the dilaton and Kalb--Ramond field respectively. Since the combination $d^2\sigma \sqrt{-h}$ is diffeomorphism invariant, the diffeomorphism invariance of $S_1$, $S_2$, and $S_3$ is manifest. Weyl invariance is manifest for $S_1$ and $S_2$ but $S_3$ in general will not be Weyl invariant. More precisely, $S_3$ is classically invariant only under global Weyl transformations. However, quantum mechanically we can force it to be locally Weyl invariant by forcing the beta function of the theory to vanish, making the theory a CFT, as discussed below. Local Weyl invariance is critical for the theory to be consistent, and so we find that classical string backgrounds correspond to solutions of the beta function equations, which in particular contain the Einstein field equations.\footnote{This reasoning has been discussed by philosophers in \citet{HuggettVistarini, Read2019-REAOMA,HuggettWuthrich}.} We'll discuss the more general situation for perturbative string theory in \S\ref{SSec: Worldsheet backgrounds}, and we'll discuss the situation for SFT in \S\ref{SSec: Backgrounds in SFT}.

\subsection{Some details about the worldsheet}\label{sec:worldsheet}

We now review some details about worldsheet string theory---i.e.,\ string theory understood as a 2D CFT---which will be relevant for the construction of covariant string field theories in the next section. 

\subsubsection{The Hilbert space}\label{Ssec: Worldsheet Hilbert Space}

Our starting point will again be the Polyakov action in flat spacetime, \eqref{eqn Polyakov}. 
When viewed as a two dimensional theory on $\Sigma_g$, the symmetries (\ref{eqn invariance}) are treated as gauge symmetries, and a gauge fixing procedure must be implemented.\footnote{See \citet{Polchinksi} for a detailed overview of the Faddeev--Popov gauge fixing for this theory.} An important consequence of the gauge fixing is the introduction of anticommuting Faddeev--Popov \textit{ghost fields} $b^{ab}$ and $c_a$, with action
\begin{equation}
    S_{\text{gh}}=\frac 1 {2\pi}\int _{\Sigma_g}d^2\sigma \sqrt{-h} b^{ab}\nabla _a c_b. 
\end{equation} 
In the quantization, one finds that the symmetries (\ref{eqn invariance}) will be anomalous unless $D=26$. Since these are local symmetries, the theory is consistent only if $D=26$. Furthermore, the gauge fixing leaves behind a residual symmetry group called the \textit{conformal group}, which is an extension of the Poincaré group that includes dilatations (scale transformations). This makes the theory a \textit{conformal field theory} or CFT.

The gauge fixing reduces the naïve Hilbert space of the theory (which is a product of the Polyakov and ghost Hilbert spaces) down to the gauge invariant states:
\begin{equation}
    \mathcal{H}_0=\mathcal{H}_{\text{P}}\otimes \mathcal{H}_{\text{gh}}\rightarrow \mathcal{H}_{\text{phys}}.
\end{equation}
This reduction is implemented by introducing an operator $Q_B$, called the \textit{BRST operator}, which acts on $\mathcal{H}_0$. The BRST operator is nilpotent ($Q_B^2=0$), and the cohomology classes of this operator give the states in $\mathcal{H}_{\text{phys}}$. These are the states in $\mathcal{H}_0$ that satisfy 
\begin{equation}\label{eqn on-shell condition}
    Q_B\ket{\psi}=0,
\end{equation}
with the equivalence
\begin{equation} \label{eqn gauge equivalence}
    \ket{\psi}\sim \ket{\psi}+Q_B\ket{\lambda}, \;\ket{\lambda}\in \mathcal{H}_0.
\end{equation}
It will be easier to work with the larger Hilbert space $\mathcal{H}_0$ while keeping in mind the equivalence (\ref{eqn gauge equivalence}). 

The states of the string are related to excitations of certain fields in spacetime. In particular, the Hilbert space $\mathcal{H}_0$ has a basis called the \textit{Fock space basis} which makes clear the relation between worldsheet states and spacetime fields. The basis is formed by acting on the vacuum state $\ket{0}$ with worldsheet creation operators $\alpha^\mu_{-n}, b_{-n}$, and $c_{-n}$, which are the modes of the fields $X^\mu, b^{ab}$, and $c_a$; the subscript $-n$ denotes the eigenvalue under the (holomorphic) worldsheet dilatation operator $L_0$ and is called the (left) \textit{level}. For the remainder of this section, we'll ignore the ghost part of the Hilbert space for simplicity. Note that there are two sets of each of the oscillators: $\alpha^\mu_{-n}$ and $\bar \alpha^\mu_{-n}$. This is because the closed string theory decouples into left-moving (holomorphic) and right-moving (antiholomorphic) sectors. The sectors are related by the \textit{level-matching condition}, which says that the left and right levels of states must be the same.\footnote{In terms of operators, $(\bar L_0-L_0)\ket{\psi}\equiv L_0^-\ket{\psi}=0$. There is also a ghost constraint $(\bar b_0-b_0)\ket{\psi}\equiv b_0^-\ket{\psi}=0$.}$^,$\footnote{Open strings are given by Riemann surfaces with boundary, and the boundary conditions couple the two sectors completely, meaning there is only one sector. Thus, there is no level-matching condition, and one can have states like $A_\mu(k)e^{ik_\nu X^\nu}\alpha^\mu_{-1}\ket{0}$, which give gauge fields in spacetime.} The vacuum state is degenerate due to the spacetime translation invariance, and the other vacua are `spacetime boosts' of $\ket{0}$:
\begin{equation}
    \ket{k}=e^{ik_{\mu} X^{\mu}}\ket{0}.
\end{equation}
The Fock space basis is an infinite-dimensional basis given by states of the form
\begin{equation}
    \alpha_{-n_1}^{\mu_1}\dots \alpha_{-n_m}^{\mu_m}\bar\alpha_{-\bar n_1}^{\mu_{m+1}}\dots \bar\alpha _{-\bar n_l}^{\mu_{m+l}}\ket{k},
\end{equation}
where $\sum n_i=\sum \bar n_i$ due to the level-matching condition.

The basis state $\alpha^{\mu_1}_{-n_1}\dots \bar \alpha^{\mu_N}_{-\bar n_N}\ket{k}$ (note the slightly different notation from above) looks like a rank $N$ spacetime tensor, since it has $N$ Lorentz indices. It will typically decompose into different irreducible representations of the spacetime Lorentz group $SO(25, 1)$, which act as a global symmetry on the worldsheet CFT: 
\begin{equation}
\alpha^{\mu_1}_{-n_1}\dots \bar\alpha^{\mu_N}_{-\bar n_N}\ket{k}=\bigoplus_\text{irreps} (\alpha^{\mu_1}_{-n_1}\dots \bar\alpha^{\mu_N}_{-\bar n_N})_{\text{irrep}}\ket{k}.
\end{equation}
For instance, the two-index state $\alpha_{-1}^\mu \bar \alpha_{-1}^\nu\ket{0}$ decomposes into a traceless symmetric, antisymmetric, and singlet representation. The component of a general state $\ket{\psi}$ along the basis vector $ (\alpha^{\mu_1}_{-n_1}\dots \bar\alpha^{\mu_k}_{-n_k})_{\text{irrep}}\ket{0}$ is interpreted as an excitation of a spacetime field which transforms in that representation.\footnote{In the more general case where some other symmetry group $G$ acts on the Hilbert space, we would also distinguish Fock space states based on their representations under that group.} For the previous example, the traceless symmetric component of $\ket{\psi}=\zeta_{\mu\nu}\alpha^\mu_{-1}\bar \alpha^\nu_{-1}\ket{0}$ is $\frac 1 2 (\zeta_{\mu\nu}+\zeta_{\nu\mu})-\zeta^\sigma _\sigma \eta_{\mu\nu} \equiv G_{\mu\nu}$ (the graviton), the antisymmetric component is $\frac 1 2 (\zeta_{\mu\nu}-\zeta_{\nu\mu})\equiv B_{\mu\nu}$ (the Kalb--Ramond field), and the singlet component is $\zeta^\sigma_\sigma \eta_{\mu\nu} \equiv \Phi \eta_{\mu\nu}$ (the dilaton). A general state can then be written in the Fock space basis as 
\begin{equation} \label{eqn Hilbert space state}
\ket{\psi} =\int \frac{d^{26}k}{(2\pi)^{26}}\left[ T(k)+(G_{\mu\nu}(k)+B_{\mu\nu}(k)+\Phi(k)\eta_{\mu\nu})\alpha^\mu_{-1 } \bar \alpha^\nu_{-1} +\dots\right]\ket{k}.
\end{equation}

The components in this basis $T(k), G_{\mu\nu}(k), \dots$ are interpreted as fields in momentum space. In position space, the graviton, for example, is given by 
\begin{align}\label{eqn: graviton fluctuation}
    G_{\mu\nu}(x)&=\int \frac{d^{26}k}{(2\pi)^{26}} e^{ikx}G_{\mu\nu}(k), \\
    g_{\mu\nu}&=\eta_{\mu\nu}+G_{\mu\nu}(x).
\end{align}
The last line is meant to illustrate that the spacetime metric \textit{in this state} is not $\eta$, but rather $\eta + G$. This is analogous to spin-2 theories of gravity, as will be discussed in \S\ref{Sec BI}. Thus, the states of the worldsheet theory are related to excitations of fields in spacetime. 

In this basis, it is easier to see what the gauge equivalence (\ref{eqn gauge equivalence}) means. The operator $Q_B$ gives rise to factors of $k_\mu$ when acting on states, and in position space this turns into a derivative $\partial_\mu$. Then, for example, the graviton component $G_{\mu\nu}$ inherits a spacetime gauge invariance\footnote{One-index spacetime states in the closed string theory arise from ghost factors in the gauge parameter state $\ket{\lambda}$, which we suppressed in the above Fock space expansion.}
\begin{equation}
    G_{\mu\nu} \sim G_{\mu\nu}+\partial_{\mu} \xi_\nu+\partial_\nu \xi_\mu,
\end{equation}
which is an infinitesimal spacetime diffeomorphism generated by the vector field $\xi_\mu$. Thus, the gauge invariance (\ref{eqn gauge equivalence}) gives rise to all gauge invariances of all spacetime fields. The condition $Q_B\ket{\psi}=0$ gives the mass-shell condition in spacetime, $k^2=-M^2$ for the various fields. The mass $M$ is related to the level by $M^2=\frac 1 {l_s^2}(n-1)$, where $l_s$ is the string length; thus, states of higher level have higher mass. Since the graviton has $n=1$, it is massless. Unfortunately, in the bosonic theory, the $n=0$ state  given by the field $T$ is tachyonic: $M^2=-\frac 1 {l_s^2}$, which suggests that the theory is unstable. Some implications of tachyons in the spectrum will be discussed in \S\ref{Section: Invariant Structure}, though this will be for \textit{open} string tachyons, which are much better understood. 

\subsubsection{Interactions}
The \textit{operator-state correspondence} is a special relation in 2D CFT between the Hilbert space of the theory and the set of operators acting on that Hilbert space. In effect, for every state, we have a local operator called the \textit{vertex operator}, and vice versa. If we wish to calculate a scattering amplitude between two asymptotic states $\ket{\psi}$ and $\ket{\phi}$, we can instead find the associated vertex operators $O_{\psi}$ and $O_{\phi}$ and compute the correlation function $\langle O_\psi O_\phi\rangle$. When the state is a Fock space basis vector $(\alpha^{\mu_1}_{-n_1}\dots \bar \alpha^{\mu_i}_{-n_i})_{\text{irrep}}e^{ik_\mu X^\mu}\ket{0}$, the associated vertex operator will be denoted $V_j(k,\sigma)$, where $k$ is the momentum of the state, $j$ is a general index specifying the irrep, and $\sigma$ is the point on the worldsheet where the operator is inserted. This vertex operator can be thought of as producing a momentum $k$ `1-particle' excitation of the field associated to that representation. 

The S-matrix between asymptotic states is therefore related to some correlation function of vertex operators. The correlation function is computed using the gauge-fixed Polyakov path integral with local vertex operator insertions, which can be viewed as summing over all intermediate worldsheet geometries with fixed asymptotic states. This is akin to summing over Feynman diagrams with fixed external states. Crucially, in string theory \textit{there is no freedom in choosing the interactions}, unlike in QFT; the Polyakov action already fixes all string interactions. 

One must be careful to preserve the gauge symmetry of the path integral. In particular, to preserve 2D diffeomorphism invariance, the vertex operators must be integrated over the entire worldsheet. Furthermore, to preserve Weyl invariance, the states associated to the vertex operators \textit{must be in the BRST cohomology}. Hence, in the worldsheet formulation of string theory, one can only compute interactions between on-shell states with $k^2=-M^2$, which means that it is difficult to compute off-shell quantities such as potentials for fields. This situation will change once we move to SFT.

There is one further technical interjection that will later prove important in the formulation of SFT. The Polyakov path integral requires an integration over the worldsheet metric $h_{ab}$ on the Riemann surface $\Sigma_g$. Due to the symmetries (\ref{eqn invariance}) and properties of 2D metrics, this infinite-dimensional integral is \textit{almost} trivial. 
That said, there are some global technicalities due to `modular parameters', which are deformations of the metric that are neither diffeomorphisms nor Weyl transformations. These parameters can be thought of as a `change of shape' of the Riemann surface; for example, on the torus $T^2=S^1 \times S^1$, these transformations roughly change the relative size/rotation of the two circles. The number of modular parameters is 0 for the sphere, 2 for the torus, and $6g-6$ for a Riemann surface of genus $g \ge 2$. We can introduce a 0, 2, or $6g-6$ dimensional space $\mathcal{M}_{g}$ called the \textit{moduli space of Riemann surfaces of genus} $g$. This space is a manifold with singularities (i.e.,\ an orbifold), and points in this space represent Riemann surfaces with particular values for the modular parameters. The path integral over metrics then reduces to an integral over modular parameters, which is an integral over this \textit{finite} dimensional space. There is a measure over this space defined in terms of the $b$-ghost, and the path integral naturally accounts for this.\footnote{See e.g.\ \citet{Nakahara}.} There is also a subtlety due to $\text{Diff} \times \text{Weyl}$ transformations that leave the metric invariant. These are called \emph{conformal Killing vectors} (CKVs); the sphere has 6, the torus has 2, and higher genus Riemann surfaces have 0. CKVs can be accounted for by inserting factors of $c$ into correlation functions.

In its full glory, the bosonic string S-matrix for external states (i.e.\ vertex operators) $V_i$ with momenta $k_i$ with the subtleties of the modular parameters (and CKVs) accounted for, is
\begin{align} \label{String S-matrix}
    S_{j_1...j_n}(k_1,\dots,k_n)&=\sum_g \int_{\mathcal{M}_g} \frac{d^{\text{dim}(\mathcal{M}_g)} t}{n_R} \int \mathcal{D} X \mathcal{D} b \mathcal{D} c \, e^{-S_P-S_\text{gh}-\lambda \chi} \nonumber \\
    \times \prod_{(a,i)\not \in f} &\int d\sigma_i^a \prod_{k=1}^{\dim{(\mathcal{M}_g)}} \frac 1 {4\pi} (b, \partial_k \hat h (t)) \prod_{(a,i)\in f} c^a(\hat {\sigma} _i) \prod_{i=1}^n \hat{h} (\sigma_i, t)^{1/2} {V}_{j_i}(k_i, \sigma_i).
\end{align}
The details needn't detain us. The rough form is just a correlation function of the vertex operators with various ghost insertions
\begin{equation} \label{rough S-matrix}
    S \sim \sum_g e^{-\lambda \chi}\int_{\mathcal{M}_g} d^{\text{dim}(\mathcal{M}_g)}t \int_{\Sigma_g} d^2\sigma \left< (b,\partial_t h)\dots c\dots V_1 \dots  V_n\right>_{\Sigma_g}.
\end{equation}
The vertex operators are integrated over the Riemann surface, the correlation function is integrated over all the moduli, and there is a sum over the genus, which can be interpreted as a loop expansion as in QFT. The different genus contributions are weighted by the Euler characteristic $\chi=2-2g$; thus, the quantity $e^{\lambda}$ acts like a coupling constant, and will be denoted $g_s$. This term arises because in 2D, the worldsheet Einstein--Hilbert action $S=\frac{\lambda}{4\pi} \int d^2 z \sqrt{h} \mathcal{R}$ is also $\text{Diff} \times \text{Weyl}$ invariant, so we should also include it in our theory in addition to the Polyakov action. This term is topological due to the Gauss--Bonnet theorem, and it is proportional to the Euler characteristic $\chi$. It has no effect other than weighting the different genus terms with different powers of $g_s$. As we can see from (\ref{S_3}), this corresponds to a constant background dilaton field $\Phi_0=\lambda\rightarrow  g_s=e^{\Phi_0}$. Thus, in string theory, the coupling is determined \textit{dynamically} by the value of the dilaton. It is not a free parameter.

\subsubsection{Backgrounds in worldsheet string theory}\label{SSec: Worldsheet backgrounds}

Much of the above discussion was in the context of the Polyakov action \eqref{eqn Polyakov} formulated around flat, 26D Minkowski space. Suppose we instead want to study bosonic string theory in a spacetime with many background fields of all spins, including a background metric field. In other words, suppose that we want to consider dynamical possibilities with objects $\langle M, \hat T, \hat g_{\mu\nu}, \hat B_{\mu\nu}, \hat \Phi, 
\ldots \rangle$.\footnote{From this point onwards, we'll denote background fields with hats; to be clear, this should not be taken to imply that there is anything quantum mechanical about these objects.} This can be done in two different ways:
\begin{enumerate}
    \item We can write down a new worldsheet action as in \S\ref{SSec StringsinBG}
    \begin{flalign} \label{eqn general nl sigma model}
        \nonumber S_P' [X^\mu, h_{ab}|\hat T, \hat g, \hat B, \hat {\Phi},\dots ] :=  \frac 1{4\pi} \int_{\Sigma_g}&d^2\sigma \sqrt{h} 
       \biggl [ \hat T(X) +\mathcal{R}\hat \Phi(X) \\ &+\Bigl ( \hat g_{\mu\nu}(X)h^{ab} +\hat B_{\mu\nu}(X)\epsilon^{ab}\Bigr ) \partial_a X^\mu \partial_b X^\nu+\dots \biggr ].
    \end{flalign}
    This theory can be quantized via a path integral. There will be a Weyl anomaly unless the couplings $\hat T, \hat g_{\mu\nu}, \hat B_{\mu\nu}, \hat \Phi, \dots$ satisfy certain field equations, including, most notably, the Einstein field equations.\footnote{See \citet{HuggettVistarini, Read2019-REAOMA} for philosophical discussion.}

    \item Instead, we might notice that inserting an operator
    \begin{equation}
        \exp\left(-\frac 1 {4\pi}\int d^2\sigma\sqrt{h}G_{\mu\nu}(X)h^{ab}\partial_aX^\mu\partial_bX^\nu\right)
    \end{equation}
    into the flat space Polyakov path integral has the effect of shifting the metric $\eta_{\mu\nu}\rightarrow \eta_{\mu\nu}+G_{\mu\nu}$. Note that this operator is $\exp(-\frac 1 {4\pi g_s}\int d^2\sigma \sqrt{h} e^{-ikX} V_{\text{grav}})$, i.e. it is the exponential of the graviton vertex operator.\footnote{Sometimes, the integral over the worldsheet $\int d^2\sigma \sqrt{h}$ is incorporated into the definition of the vertex operator.} More generally, if we want a background value for a field corresponding to the vertex operator $V_j$, we simply insert the operator $\exp(-\frac 1 {4\pi}\int d^2\sigma \sqrt{h} V_j)$ into the path integral; this is a so-called coherent state.\footnote{For philosophical discussion, see \citet{HuggettWuthrich, ReadBI}.} Again, this can only be done for BRST invariant operators $V_j$, which means this is a `marginal deformation', i.e.\ preserves conformal invariance. Thus, backgrounds obtained in this way also satisfy the field equations from above. 
\end{enumerate}

\subsubsection{Superstring theory}\label{sec: superstring theory}

In superstring theory, in addition to worldsheet conformal symmetry, we require worldsheet supersymmetry, which is a symmetry that takes bosons into fermions and vice versa. Such theories are called \textit{superconformal field theories} or SCFTs. We can impliment this symmetry by introducing worldsheet fermions into the action (written in complex worldsheet coordinates $z=\sigma +i\tau$):
\begin{equation}\label{eq:superstringaction}
    S=\frac 1 {4\pi} \int d^2 z \left ( \partial X^\mu \bar \partial X_\mu +\psi^\mu \bar \partial \psi_\mu +\tilde{\psi}^\mu \partial \tilde{\psi}_\mu \right)+S_{\text{ghost}}.
\end{equation}
Fermions can be either periodic or antiperiodic when going around the string. Periodic fermions are said to be in the R (Ramond) sector, while antiperiodic fermions are in the NS (Neveu--Schwarz) sector. For the anomaly to vanish in this theory, we find $D=10$.

The action \eqref{eq:superstringaction} is for the `type II' closed string theory, which has a holomorphic SCFT and an antiholomorphic SCFT.\footnote{The distinction between IIA and IIB superstring theories arises from a difference in \textit{GSO projection} between the two sectors, which truncates the spectrum and is needed to have a consistent theory.} It has four sectors depending on the periodicities of $\psi$ and $\tilde \psi$: NSNS, RNS, NSR, and RR. The NSNS states are familiar from bosonic string theory and include the graviton, Kalb--Ramond $B$-field, and dilaton. The RNS and NSR sectors are fermions \textit{in spacetime} and in fact give rise to an $\mathcal{N}=2$ supersymmetry in spacetime.\footnote{\textit{A priori} supersymmetry on the worldsheet and supersymmetry in spacetime have nothing to do with each other, so this is a surprising fact from the worldsheet point of view.} The RR sector has new bosonic higher-form field strengths, and these pose certain difficulties which will be discussed later. 

The heterotic strings are formed by combining a holomorphic SCFT with an antiholomorphic bosonic CFT.\footnote{There is a mismatch in central charge ($c=15$ for the holomorphic part and $\bar c=26$ for the antiholomorphic part) between the two sectors; on the bosonic side, the interpretation of a particular boson as a coordinate only holds for $10$ bosons, so $D=10$ in spacetime. The remaining $16$ bosons are interpreted as giving rise to the gauge group. For philosophical discussion of this, see \citet{Kratky-2024}.} Thus we just have an R sector and an NS sector depending on the periodicity of $\psi$, which are fermions and bosons in spacetime respectively. There is an $\mathcal{N}=1$ supersymmetry in spacetime, as well as a spacetime gauge group SO(32) or $E_8\times E_8$.

In addition to the diffeomorphisms and Weyl transformations, we need to gauge fix the local supersymmetry. This is done by introducing additional ghost fields, which again have opposite statistics to the fields to which they are associated. Since supersymmetries are generated by anti-commuting spinors, the ghosts will be commuting fields. They are denoted as the `$\beta \gamma$' CFT; the holomorphic part of the ghost action is
\begin{equation}
    S_g= \frac 1 {2\pi} \int d^2z (b\bar \partial c +\beta \bar \partial \gamma).
\end{equation}

Recall that if we wish to compute scattering amplitudes between states, we must find the appropriate vertex operators corresponding to those states. In the superstring case, there turn out to be difficulties in this process due to the definition of the vacuum. Essentially, one finds that the operator corresponding to the vacuum must satisfy very strange relations from the perspective of the $\beta \gamma$ CFT. For instance, the NS vacuum $\ket{0}_\text{NS}$ maps to the delta function $\delta(\gamma)$!

Thus, instead of using the $\beta \gamma$ ghosts, it is useful to introduce an equivalent CFT formulated in terms of an anticommuting ghost system `$\xi \eta$' (which are like the $bc$ ghosts) and a free boson $\phi$:
\begin{equation}
    \beta(z) \cong e^{-\phi(z)} \partial \xi(z), \;  \gamma \cong e^{\phi(z)} \eta(z).
\end{equation}
In terms of these variables, the vacua are easily expressed as $\ket{0}_\text{NS}\cong e^{-\phi}$ and $\ket{0}_{\text{R}}\cong e^{-\phi/2}$. Under the current $j=\partial \phi$, these operators have charge $-1$ and $-1/2$ respectively; this charge is called the \textit{picture number}. Since general physical states are constructed by acting on the vacuum with matter creation operators, general vertex operators in the NS and R sectors will also have picture number $-1$ and $-1/2$ respectively, i.e., they will contain factors of $e^{-\phi}$ and $e^{-\phi/2}$ respectively.

One must also worry about the fermionic analogues of the moduli and CKVs when computing superstring scattering amplitudes. Recall that in the bosonic case, we needed to insert factors of $b$ and $c$ to compensate for these subtelties. In the fermionic case, we must instead insert  \textit{picture changing operators} (PCOs):
\begin{equation}
    \mathcal{X}(z):= \{Q_B, \xi\}.
\end{equation}
The explicit form of the PCO is quite messy, but it is crucial in superstring scattering amplitudes and shows up in the formulation of superstring field theory. We will just use the zero mode of this operator
\begin{equation}\label{eqn: PCO}
    \mathcal{X}_0=\frac 1 {2\pi i} \oint \frac 1 z \mathcal{X}(z).
\end{equation}
As the name suggests, $\mathcal{X}_0$ adds $1$ to the picture number of the amplitude.

The superstring S-matrix has a very similar form to the bosonic case but with additional $\beta \gamma$ and $\mathcal{X}_0$ insertions. Background fields are much more subtle in the superstring case, however (see e.g.\ \citet{Backgroundsuperstring}). Background RR sector fields in particular \textit{cannot} be incorporated using a straightforward worldsheet action or the coherent state method; roughly, the reason is because R sector vertex operators are worldsheet spinors, and they cannot be exponentiated in the right way. String field theory will provide a workaround for this difficulty.

\section{String field theory}\label{Sec StringFieldT}

Perturbative string theory based upon the Polyakov action has its shortcomings. In particular, there are issues with IR divergences and renormalization,\footnote{There are no UV divergences due to the properties of moduli spaces of Riemann surfaces that control loop integrals.} as well as the aforementioned problem with background RR fields. For these reasons and others, it is therefore useful to consider a second-quantized version of the string, i.e.\ a `string field theory' (SFT).\footnote{For recent reviews of SFT, see \citet{Erbin, sen2024stringfieldtheoryreview}.} Roughly, the SFT is a special QFT whose scattering amplitudes reproduce all Polyakov scattering amplitudes (\ref{String S-matrix}). This might seem contradictory: the Polyakov scattering amplitudes include UV \textit{finite} graviton scattering amplitudes, and a standard QFT would not be able to accommodate this due to the nonrenormalizability of gravity in higher dimensions. Indeed, SFT gets around this issue with various subtleties: it has infinite fields of higher mass and spin which `tame' the UV behaviour, the action is in general nonlocal in the position space representation, and it has an intricate gauge structure. 

SFTs are based on the string field $\ket{\Psi} \in \mathcal{H}$, which is an arbitrary (Grassman even) member of the Hilbert space of the worldsheet theory satisfying the level matching condition $L_0^- \ket{\Psi}=0$.\footnote{This is needed for consistent interactions. Henceforth, $\mathcal{H}\subset \mathcal{H}_0$ will denote the subspace of the Hilbert space which meets these conditions.} It might look odd having the field written as a ket instead of an operator, but it can be translated into a more conventional field by using the Schr{\"o}dinger representation, which is a functional in spacetime $\braket{X^\mu, c| \Psi }=\Psi[X^\mu(\sigma),c]$. The string field depends on the shape of the string in spacetime (and also the $c$ ghost), not just a point like a conventional quantum field; this is the source of nonlocality in the interactions. We will also use the symbol $\Psi$ to represent the vertex operator associated to $\ket{\Psi}$.

Although the string field can be represented as a functional on loop space $\Psi[X^\mu(\sigma),c]$, if we use the Fock space basis of the Hilbert space, we can expand it as a collection of infinite \textit{point-valued} fields, which represent the various modes of the string. One defines the quantum theory by doing a path integral over $\Psi$, which amounts to doing infinite path integrals over each of the components of $\Psi$, i.e.,\ infinite path integrals over these classical fields. 


\subsection{Bosonic closed string field theory}

The kinetic part of the action for the bosonic closed string field theory has the form\footnote{The $c_0^-=c_0-\bar c_0$ ghost is part of the definition of the inner product.}
\begin{equation} \label{kinetic SFT action}
    \frac 1 2 \bra{\Psi} c_0^- Q_{B} \ket{\Psi},
\end{equation}
which tells us that on-shell states are in the BRST cohomology, 
\begin{equation}
    Q_{B}\ket{\Psi}=0.
\end{equation}
This is analogous to the Klein-Gordon part of scalar field theory: $(\square - m^2)\phi=0$, which tells us that on-shell states satisfy $-k^2=m^2$, exactly like the BRST condition. The free theory only has this kinetic term in its action, and so it has a gauge symmetry given by $\ket{\Psi} \rightarrow \ket{\Psi}+Q_{B}\ket{\Lambda}$, since $Q_B^2=0$.

To get a better picture of what this means, we can write down a spacetime action by expanding the string field in the Fock space basis of the Hilbert space (where we ignore ghost factors for simplicity):
\begin{equation}\label{eqn: Fock Space Expansion of string field}
    \ket{\Psi} = \int \frac{d^{26}k}{(2\pi)^{26}}[T(k)+(G_{\mu\nu}(k)+B_{\mu\nu}(k)+\Phi(k) \eta_{\mu\nu})\alpha_{-1}^{\mu}\bar \alpha_{-1}^{\nu}+\dots] \ket{k}.
\end{equation}
The coefficients $T(k), G(k), \dots$ are interpreted as Fourier transforms of the spacetime fields $T(x), G(x), \dots$. In position space, the action (\ref{kinetic SFT action}) gives kinetic terms for all of the fields:
\begin{equation}
    S=\int {d^{26}x}\left( \frac 1 2 \eta^{\mu \nu}\partial_\mu T(x) \partial _\nu T(x) + \dots \right),
\end{equation}
where we have assumed our worldsheet theory is 26D flat Minkowski space. The free SFT is therefore a quantum field theory with infinite fields which represent the various vibrational modes of the strings. This move to position space will not work for the action of the interacting theory. 

To write down the interacting SFT action, we need to go beyond the Polyakov S-matrix from above. We need amplitudes of off-shell vertex operators ($Q_B\ket{V}\not = 0$), which the Polyakov formalism cannot deal with (see \citet[p.\ 103]{Polchinksi}). Once these amplitudes are known, we can reverse-engineer an action whose `Feynman diagram' vertices give rise to them. 

It is therefore necessary to consider off-shell string states which are not necessarily conformally invariant, which means they will be coordinate-dependent.\footnote{This is because we need to insert factors of $\int d^2z V$ into amplitudes, and such factors will be coordinate dependent unless $V$ is a $(1,1)$ primary (i.e. $Q_B\ket{V}=0$), since the transformation of $d^2z$ will not cancel that of $V$.} This can be done by introducing local coordinates around each puncture on the Riemann surfaces where the vertex operators are inserted. We can systematically account for this using a fiber bundle with base space $\mathcal{M}_{g,n}$, the moduli space of Riemann surfaces of genus $g$ with $n$ punctures,\footnote{This space is $\text{dim}(\mathcal{M}_g)+2n$ dimensional, where the dimension of $\mathcal{M}_g$ is discussed above. This is because this space includes both the geometric moduli as well as the 2 coordinates of each of the $n$ punctures.} and with the fiber being a choice of local coordinates around each puncture (up to a phase). This bundle is denoted $\hat{\mathcal{P}}_{g,n}$.\footnote{A point in this bundle will be a Riemann surface of genus $g$ with $n$ punctures with a choice of local coordinates around each of these punctures. The projection $\pi: \hat P_{g,n}\to \mathcal{M}_{g,n}$ simply `forgets' the local coordinates. Note that the fiber itself is infinite dimensional.}

We now want to look for an amplitude constructed with this space which reproduces Polyakov amplitudes when the vertex operators are on-shell but which still works off-shell. We begin by simply picking specific local coordinates for each Riemann surface with fixed moduli and vertex operator positions, i.e.\ we pick a specific section $\mathcal{F}_{g,n}$ of $\hat{\mathcal{P}}_{g,n}$. If we integrate over $\mathcal{F}_{g,n}$, then we will also be integrating over all of the moduli and all of the vertex operator positions, since both are already accounted for in the base space $\mathcal{M}_{g,n}$. Thus, to get the genus $g$ contribution to the $n$-point amplitude, we need to integrate some $p=\text{dim}(\mathcal{M}_{g,n})$ form over this section:
\begin{equation} \label{Off-shell amplitude}
    \mathcal{A}_g(V_1,\dots ,V_n)=(g_s)^{-\chi_{g,n}}\int _{\mathcal{F}_{g,n}} \Omega^{(g,n)}_{p} (V_1,\dots ,V_n).
\end{equation}

The factor of $g_s$ plays the role of the coupling constant and appears for the same reason as in (\ref{String S-matrix}). The specific $p$-form can be determined by the condition that it reproduces (\ref{String S-matrix}) when the $V_i$ are all on-shell:

\begin{equation}
    \Omega^{(g,n)}_{p}\left[ \frac{\partial}{\partial u^{j_1}}, \dots , \frac{\partial}{\partial u^{j_p}} \right]=\left( -\frac 1 {2\pi i}\right)^{3g-3+n} \left<  \mathcal{B} \left[ \frac{\partial}{\partial u^{j_1}} \right]\dots \mathcal{B} \left[ \frac{\partial}{\partial u^{j_p}} \right] V_1 \dots V_n \right>.
\end{equation}
$\mathcal{B}\left[ \frac{\partial}{\partial u }\right]$ is an operator constructed from the $b$-ghosts and is a one-form on the tangent space of $\hat{\mathcal{P}}_{g,n}$. The total amplitude is then
\begin{equation}
    \mathcal{A}(V_1,\dots,V_n)=\sum_{g=0}^{\infty}\mathcal{A}_g(V_1,\dots,V_n).
\end{equation}

Essentially, the amplitude has the form of a normal Polyakov string amplitude (\ref{rough S-matrix}) with vertex operators and ghost insertions, just with the new detail of local coordinates, which modify the ghost insertions slightly and require the integration over the section $\mathcal{F}_{g,n}$ instead of just $\mathcal{M}_{g,n}$. The integration over $\mathcal{F}_{g,n}$ contains the integration over moduli \textit{and} vertex operator positions, so it combines the two types of integrals we saw in the S-matrix. Note that all physical quantities are independent of the choice of section, which means they are independent of the local coordinates.


We now want to reverse-engineer an action by using the amplitudes. However, we need to be careful. The difficulties are illustrated in a simple example from QFT (drawn from \citet[p.\ 295]{Erbin}). Suppose we have a $\phi^3+\phi^4$ scalar field theory. Now consider a $2\rightarrow 2$ scattering. There are tree-level contributions from the $\phi^3$ term, with two interactions connected by an internal propagator. However, since there is also a $\phi^4$ interaction, there will also be a tree-level interaction with all particles connecting at one point. If we are trying to reverse-engineer the action by just looking at amplitudes, we need to properly account for both of these effects. For example, if we know all of the lower-order interactions up to the $n$-point scattering (here the $\phi^3$ term), and we know the propagator, then we can deduce the existence of a $\phi^4$ term in the action, since the $\phi^3$ diagrams alone do not give the full amplitude.

Similarly, in SFT, the $n$-point amplitude at genus $g$ (\ref{Off-shell amplitude}) will have many contributions coming from other-order interactions with propagators. However, some parts of the amplitude will not be accounted for if we just use these interactions, and so we must add a new $n$-point interaction term in the action. This will continue for all $n$, so there will be infinite interaction terms.\footnote{An important exception is Witten's cubic open string field theory, which can reproduce all interactions with just a cubic term in the action.
This will be discussed in \S\ref{SSec: Cubic SFT}.} To describe these terms explicitly, we need to understand how the other-order interactions contribute to the $n$-point amplitudes. 

There are two important gluing operations on the moduli spaces of Riemann surfaces that allow one to construct higher genus Riemann surfaces from lower genus Riemann surfaces. In particular, starting with two punctured Riemann surfaces $\Sigma^1_{g_1,n_1}$ and $\Sigma^2_{g_2,n_2}$, we can construct a new Riemann surface $\Sigma_{g,n}$ of genus $g=g_1+g_2$ with $n=n_1+n_2-2$ punctures by gluing a puncture of $\Sigma^1$ with a puncture of $\Sigma^2$. This can be denoted by $\{\Sigma^1, \Sigma^2\}=\Sigma$. The other gluing operation just glues two punctures of the same Riemann surface $\Sigma^3_{g_3,n_3}$ which gives a new Riemann surface $\Sigma_{g,n}$ of genus $g=g_3+1$ and $n=n_3-2$ punctures. This operation can be denoted $\Delta \Sigma^3=\Sigma$.\footnote{In both of these operations, there is an arbitrary complex parameter which determines exactly how the gluing is performed. This is discussed in more detail in the next footnote.}

These two operations can be interpreted as follows. The operation of gluing two distinct Riemann surfaces should be understood as two separate Feynman diagrams with a propagator running between them. In other words, this is a 1PR (one particle reducible) graph with all interactions coming from lower-order vertices.
Similarly, the operation of gluing a surface to itself gives the contribution coming from higher point vertices with internal loops. This latter effect did not show up in the tree-level $\phi^3+\phi^4$ example from above, since it arises from loops.  

Now, by the logic above, in order to find the $n$-point interaction term in the action, we need to look at the part of (\ref{Off-shell amplitude}) that is not accounted for by other-order interactions that are glued together. If we glue all possible Riemann surfaces of appropriate $g_i$ and $n_i$ in all possible ways, we will cover a region of $\mathcal{M}_{g,n}$.\footnote{\label{Stubs}We might worry about covering the same region twice, which would over-count certain contributions. This can be avoided by adjusting a parameter in the gluing operation called the \textit{stub parameter}. Although the stub parameter can change the form of the string vertices defined below, it can be shown that SFTs with different stub parameters are related by field redefinitions and are hence equivalent.} However, there will be some subset of $\mathcal{M}_{g,n}$ that is not covered. This in turn will correspond to some subset of the section $\mathcal{F}_{g,n}$; it is denoted $\mathcal{V}_{g,n}$ and is called a \textit{string vertex.} In the language of the QFT analogy, this string vertex is akin to the $\phi^4$ interaction. It must appear in the action because the other interaction terms are not able to reproduce the full amplitude by themselves. As one might expect, these vertices must satisfy complicated and strict geometric relations so as to perfectly cover all moduli spaces of Riemann surfaces. This will give rise to a geometric BV master equation, which is discussed in \S\ref{SSec: BV}.

Using the string vertices $\mathcal{V}_{g,n}$, we can define a multilinear bracket $\{A_1, \dots,A_n\}: \mathcal{H}^{\otimes n} \rightarrow \mathbb{C}$,
\begin{equation}
    \{ A_1,\dots ,A_n\}:=\sum_{g=0}^{\infty}(g_s)^{-\chi_{g,n}}\int_{\mathcal{V}_{g,n}} \Omega^{(g,n)}_{\text{dim}(\mathcal{M}_{g,n})}(A_1,\dots ,A_n).
\end{equation}
The action of the interacting quantum bosonic closed string field theory is then 
\begin{equation} \label{eqn: bosonic SFT action}
    S=\frac 1 2 \bra{\Psi} c_0^- Q_B \ket{\Psi} + \sum_{n=1}^\infty \frac 1 {n!}\{\Psi^n\},
\end{equation}
where
\begin{equation}
\{\Psi^n\}:=\underbrace{\{ \Psi,\dots ,\Psi\}}_n.
\end{equation}
Due to the geometric properties of the vertices, this action will be able to reproduce all bosonic closed string Polyakov amplitudes. 

This action has a complicated gauge invariance: 
\begin{equation}\label{eqn: closed bosonic gauge invariance}
\ket{\Psi}\rightarrow \ket{\Psi}+Q_B \ket{\Lambda}+\sum_{n=1}^\infty \frac 1 {n!} \ket{[\Lambda \Psi^n]},
\end{equation}
where
\begin{equation} \label{eqn: string bracket}
    \bra{A_0} c_0^- \ket{[A_1,\dots ,A_n]} :=\{A_0,\dots ,A_n\}
\end{equation}
defines the \textit{string bracket}: $\ket{[A_1,\dots,A_n]}: \mathcal{H}^{\otimes n} \to \mathcal{H}$. A common choice to gauge fix this symmetry is $b_0^-\ket{\Psi}=0$, which is called \textit{Siegel gauge}. The classical equation of motion is
\begin{equation}
    Q_B \ket{\Psi }+ \sum_{n=1}^\infty \frac 1 {n!} \ket{[\Psi^n]} = 0.
\end{equation}
Quantum observables will be discussed in \S\ref{SSec: BV}, since the gauge invariance (\ref{eqn: closed bosonic gauge invariance}) will restrict the space of physical operators. 

\subsection{Closed superstring field theory}
Superstring field theory is structurally very similar to bosonic string field theory. One again has a physical string field $\Psi$, now defined to have picture number $-1$ in the NS sector and $-1/2$ in the R sector. The reason for these assignments is the same as discussed in \S\ref{sec: superstring theory}. 

A major difference in the superstring case, however, is the introduction of an auxiliary, unphysical string field $\tilde \Psi$, which is defined to have picture number $-1$ in the NS sector and $-3/2$ in the R sector. This field is needed to write down a kinetic term for the R sector. As a motivation for its introduction, one might recall the presence of the self-dual RR $5$-form in type IIB supergravity/string theory. The difficulties in writing an action for this field are well known. Since this field should arise within the type II superstring field theory, similar issues arise here. The auxiliary field, however, provides a workaround for this difficulty.\footnote{Another difference from the bosonic case is that the section $\mathcal{F}$ has to be replaced with a more general object called a \emph{chain}, which must appropriately project down onto $\mathcal{M}$. The reason is that picture changing operators can collide with each other and with vertex operators, which lead to \textit{spurious divergences}. A technique called \textit{vertical integration} is needed to avoid these collisions, and it requires vertical segments on $\mathcal{F}$, which means it is no longer a section.} 

Define $\mathcal{G}$ to be 1 in the NS sector and $\mathcal{X}_0$, the zero mode of the picture changing operator (\ref{eqn: PCO}), in the R sector. Then the closed superstring field theory action is 
\begin{equation}\label{eqn: superstring field theory action}
    S=-\frac 1 2 \bra{\tilde \Psi} c_0^- \mathcal{G} Q_B \ket{\tilde \Psi}+ \bra{\tilde \Psi} c_0^- Q_B \ket{\Psi} +\sum_{n=1}^\infty \frac 1 {n!} \{ \Psi^n\},
\end{equation}
and the equations of motion become
\begin{align}
    &Q_B\ket{\Psi} -\mathcal{G} Q_B\ket{\tilde \Psi}=0 \\
    &Q_B \ket{\tilde \Psi} + \sum_{n=1}^\infty \frac 1 {n!}\ket{[\Psi^n]}=0,
\end{align}
which simplify to 
\begin{align}
    &Q_B\ket{\Psi}+\sum_{n=1}^\infty \frac 1 {n!}\mathcal{G} \ket{[\Psi^n]}=0. \label{supereom} \\
    &Q_B\ket{\Psi} = \mathcal{G} Q_B\ket{\tilde \Psi}. 
\end{align}
One can solve these equations by finding $\Psi$ using the first equation and $\tilde \Psi$ with the second. We will not write $\tilde \Psi$ explicitly when describing solutions of the superstring field theory, since it is completely determined by $\Psi$. This action also has a gauge invariance:
\begin{align}\label{eqn: closed superstring gauge invariance}
    &\ket{\Psi} \rightarrow \ket{\Psi}+Q_B \ket{\Lambda} +\sum_{n=2}^\infty \frac 1 {n!}\mathcal{G}\ket{[\Lambda  \Psi ^n]}\\
    &\ket{\tilde \Psi} \rightarrow \ket{\tilde \Psi} + Q_B\ket{\tilde \Lambda} + \sum_{n=2}^\infty \frac 1 {n!}\ket{[\Lambda  \Psi ^n]}.
\end{align}

\subsection{BV quantization and $L_\infty$ algebras}\label{SSec: BV}

As we have seen, the string field theory action has a gauge symmetry given by (\ref{eqn: closed bosonic gauge invariance}) in the bosonic case and (\ref{eqn: closed superstring gauge invariance}) in the supersymmetric case. To compute physical observables and ensure consistency, one must identify the gauge--invariant operators and verify that the quantum theory is anomaly free. The BV formalism is a powerful quantization scheme which allows one to do both of these things. We provide here a brief sketch of the essential points.\footnote{See e.g.\ \citet{BV} for more details.}


When we decomposed the moduli spaces of Riemann surfaces into various vertices, we needed to ensure that these vertices perfectly covered the moduli spaces without any gaps or overlaps so that we did not double count any contributions to the scattering amplitudes. This consistency condition can be expressed as a \textit{geometric BV master equation}:
\begin{equation}\label{eqn: geometric BV}
    -\partial \mathcal{V}_{g, n}=\Delta \mathcal{V}_{g-1, n+2}+\frac 1 2   \sum_{\substack{g_1+g_2=g \\ n_1+n_2-2=n}} \{\mathcal{V}_{g_1, n_1}, \mathcal{V}_{g_2,n_2}\}.
\end{equation}
To explain the notation, recall that $\mathcal{V}_{g,n}$ is a section over a subspace of $\mathcal{M}_{g,n}$. $\partial \mathcal{V}_{g,n}$ is defined as the boundary of the projection of this subspace down to $\mathcal{M}_{g,n}$, i.e. the boundary of $\pi(\mathcal{V}_{g,n})$. The operations $\Delta \mathcal{V}$ and $\{\mathcal{V}_1, \mathcal{V}_2\}$ are defined as the spaces formed by performing the operations $\Delta\Sigma$ and $\{\Sigma_1, \Sigma_2\}$ for all $\Sigma \in \pi(\mathcal{V})$ and $\Sigma_1\in \pi(\mathcal{V}_1), \Sigma_2\in \pi(\mathcal{V}_2)$ respectively. In other words, similarly to above, we glue all Riemann surfaces in the vertices of appropriate $g,n$ using the operations $\Delta A$ and $\{A ,B\}$. Addition means union, and the negative indicates that the orientations are opposite. 

The above equation is so-named because of its algebraic relation to another equation called the \textit{BV master equation}:
\begin{equation}\label{eqn: BV master equation}
    \Delta S + \frac 1 2  \{S, S\} =0.
\end{equation}
This equation has its origins in the BV quantization of gauge theories, which is the most powerful technique for quantizing theories with intricate gauge symmetries. BV quantization is extremely powerful for understanding string field theory and its gauge-invariant observables, so let us quickly review how it works. 
 
Suppose we want to quantize a classical action $S_0$ which has some gauge symmetry $G$, and perhaps even gauge symmetries of gauge symmetries $G'$, and so on.\footnote{This happens in higher-form theories. Suppose we have a theory based on an abelian two form potential $B$ whose field strength is $H=dB$. The potential $B$ has a gauge transformation under $B\to B+d\Lambda$, where $\Lambda$ is a one form. However, $\Lambda$ also has a gauge transformation under $\Lambda \to \Lambda +d\lambda $, where $\lambda$ is a zero form. In string field theory, this process continues infinitely!} One introduces ghosts for each generator of the gauge group (and ghost of ghosts, and so on). Then, the set of physical fields and ghosts is collectively called the set of \textit{fields}, denoted $\psi^a$. For every field, we now introduce an \textit{antifield} $\tilde \psi ^a$. Suppose we have $N$ fields, i.e. $a:1, \dots, N$. We can then combine the fields and antifields into a $2N$ dimensional (super)manifold $M$ whose coordinates are given by $(\psi ^a, \tilde \psi ^a)$. We will use a combined index $I:1, \dots, 2N$ such that for $I: 1, \dots, N$, $\psi ^I = \psi ^a$, and for $I: N+1, \dots, 2N$, $\psi ^I=\tilde \psi ^a$. Introduce a sympectic form on this space called $\omega^{IJ}$.\footnote{This is the inverse matrix of the components of the symplectic form $\omega_{IJ}$ in the coordinates $\psi^I$.} Using this form, we can define the \textit{BV antibracket} between functions on $M$: 
 \begin{equation}\label{eqn: BV antibracket}
     \{F, G\} = \frac{\partial_r F}{\partial \psi ^I} \omega^{IJ}  \frac{\partial_l G}{\partial \psi ^J}, 
 \end{equation}
and we can define the \textit{BV Laplacian}, 
\begin{equation}\label{eqn: BV Laplacian}
    \Delta F = \frac 1 2(-1)^{\psi^I} \frac{\partial _l}{\partial \psi ^I} \left( \omega^{IJ} \frac{\partial_r F}{\partial \psi ^J} \right).
\end{equation}
Roughly, the antibracket `generates' classical gauge transformations, just like the Poisson bracket generates time translations: $\delta A = \{S, A\}$. The Laplacian $\Delta$ gives quantum corrections to this relation coming from the path integral measure.\footnote{The subscripts $r$ and $l$ and the exponent $(-1)^{\psi ^I}$ have to do with the Grassmanality of the (anti)fields, which we will not discuss for simplicity.}  These operations can be used to identify the gauge invariant observables, as will be seen in (\ref{eqn: gauge invariant observables}). In string field theory, they are related to the operations on the moduli spaces of Riemann surfaces which we introduced above.

To quantize the theory $S_0$, we first extend it to a function $S$ on $M$ that reduces back to $S_0$ when the antifields are set to zero: 
\begin{equation}
    S_0(\psi ^a) \to S(\psi^a, \tilde \psi^a):  S(\psi ^a, 0) = S_0(\psi^a).
\end{equation}
If this extended action $S$, called the \textit{quantum master action}, satisfies the above BV master equation (\ref{eqn: BV master equation}), then it will define an anomaly-free, gauge-invariant quantum theory. Gauge fixing involves choosing an arbitrary (odd) function $F$ on $M$ and imposing the condition $\tilde \psi^a = \frac{\partial_l F }{\partial \psi ^a} $. Then, the quantum theory is based on the gauge fixed path integral\footnote{Note the unusual sign in the exponent.}
\begin{equation}
    \int \mathcal \prod_a d\psi^a e^S.
\end{equation}
The gauge-invariant observables $A$ of this quantum theory are functions on $M$ that satisfy 
\begin{equation}\label{eqn: gauge invariant observables}
    \Delta A+\{ S, A\} = 0,
\end{equation}
and their expectation values are determined by 
\begin{equation}
    \langle A \rangle = \int_L \mathcal{D}\psi ^a A(\psi ^a)e^S,
\end{equation}
where $L$ is a Lagrangian submanifold of the full supermanifold $M$.\footnote{A Lagrangian submanifold is a submanifold where the symplectic form vanishes, $\omega=0.$}

Indeed, the SFT action (\ref{eqn: bosonic SFT action}) satisfies the master equation (\ref{eqn: BV master equation}): the space of field configurations $M$ is just the Hilbert space $\mathcal{H}$ of the worldsheet theory, the antibracket is determined in terms of the CFT inner product, and the split into fields and antifields is based on the worldsheet ghost number of the state of the CFT. Notice that this is a somewhat backwards story to the one which we told above: there we had a classical action we were trying to quantize, and we had to introduce ghosts, antifields, and other objects. In SFT, the entire BV structure is already there, and we already have a gauge-fixed master action (when we impose Siegel gauge). It follows that SFT is a consistent, anomaly-free, gauge-invariant quantum theory. The quantum observables of string field theory are functions $F(\Psi)$ satisfying 
\begin{equation} \label{eqn: gauge invariant observable condition}
    \Delta F+\{ S, F\}=0,
\end{equation}
and their expectation values are given by
\begin{equation}\label{eqn: gauge invariant VEV}
    \langle F \rangle =\int_L \mathcal{D} \Psi F(\Psi) e^{S(\Psi)}.
\end{equation}

The algebra satisfied by $\{,\}$ and $\Delta$ (potentially including more operations given by brackets $[\cdot]$) is called a (generalized) \emph{BV algebra}. Then, in more abstract terms, the SFT is determined by a morphism from the BV algebra of the vertices, which is defined by the decomposition of the moduli spaces of Riemann surfaces, into the BV algebra of multilinear functions on the Hilbert space of the CFT, which is defined by the string brackets. Due to the properties of the BRST charge and its action on vertex operators, these brackets satisfy the axioms of a \textit{quantum} $L_\infty$ \textit{algebra}.\footnote{See \citet{sen2024stringfieldtheoryreview} for a detailed introduction to $L_\infty$ (and $A_\infty$) structures.}  

One can show that different SFTs based on the same background, i.e.\ different SFTs based on the same worldsheet theory with Hilbert space $\mathcal{H}$, are related by $L_\infty$ isomorphisms \citep{muenster2012homotopyclassificationbosonicstring, Kajiura}. Moreover, such isomorphisms preserve the action of the theory. Thus, there is a good case that $L_\infty$ isomorphism is the correct standard for the equivalence of SFTs. Since all string field theories formulated using the above construction will end up with the same $L_\infty$ structure, but merely with different definitions for the string vertices and sections, we see that there is a unique SFT corresponding to every worldsheet background, and the various choices made in formulating the theory are irrelevant.

Concretely, the different $L_\infty$ isomorphisms are field redefinitions of the string field that preserve the BV structure and action. In \S\ref{Ssec: schema for BI}, we will focus on background independence proofs that use this latter (more familiar) notion of equivalence, though the $L_\infty$ approach has also produced results pointing in similar directions \citep{muenster2012homotopyclassificationbosonicstring}.

\subsection{The 1PI action}

Similar to standard QFT, we can write down a 1PI (one particle irreducible) effective action for SFT. Recall that such an action is defined as the Legendre transformation of the log of the partition function:
\begin{equation}\label{eqn: 1PI}
    \Gamma_{\text{1PI}}[\phi]=W[J]-\int d^4 x J(x)\phi(x),
\end{equation}
and $W[J]=-i\ln Z[J]$. Often, classical backgrounds of a QFT (meaning solutions to the classical equations of motion) get corrected by loops and can drastically change behaviour. However, the 1PI action takes all of these corrections into account, and solutions to the equations of motion of the effective action tell us the good quantum vacua. Note that the 1PI action is treated as a classical action, meaning it satisfies the classical BV master equation $\{S, S\}=0$ rather than the quantum master equation (\ref{eqn: BV master equation}).


The 1PI action generates, as its name suggests, all 1PI Feynman diagrams.
It can be constructed in SFT by simply changing our definition of the string vertices: instead of letting $\mathcal{V}_{g,n}$ be a section over the region of $\mathcal{M}_{g,n}$ not covered by $\{\Sigma_1, \Sigma_2\}$ and $\Delta \Sigma_3$, we will let it be a section over the region not covered by $\{\Sigma_1 , \Sigma_2\}$ only. In other words, the loop contributions from $\Delta \Sigma_3$ are being absorbed into the string vertices, just leaving separating Riemann surfaces connected by propagators, i.e.\ the 1PR graphs. Since the only change is the definition of $\mathcal{V}_{g,n}$, the form of the action is exactly the same, just replacing $\{A_1,\dots ,A_n\}$ with $\{A_1,\dots ,A_n\}_{\text{1PI}}$, which uses the new vertices. Since we will mainly use the 1PI action, we will drop the 1PI subscript. As we will see later in \S\ref{Ssec: schema for BI}, the difficulties introduced by $\tilde \Psi$ in the superstring field theory mean that it is much easier to prove background independence using this effective action and its equation of motion.

\subsection{Backgrounds in string field theory}\label{SSec: Backgrounds in SFT}

The 1PI equation of motion for the closed superstring field theory takes the same form as the classical  equation of motion, but it uses the 1PI vertex in defining the string bracket $\ket{[A_1, \dots, A_n]}$: 
\begin{equation} \label{EOM}
     Q_B \ket{\Psi}+\sum_{n= 1}^\infty \frac 1 {n!} \mathcal{G} \ket{[\Psi^n]}=0.
\end{equation}
As mentioned above, solutions to this equation are the backgrounds of the full quantum SFT. The simplest solution is $\Psi = 0$; this is the solution representing the CFT which we used in constructing the SFT. 

Hence, there are two different ways of understanding the backgrounds of string theory:
\begin{enumerate}
    \item A background is given by the worldsheet CFT, which in turn corresponds to some set of background fields in the spacetime, as in (\ref{eqn general nl sigma model}). Since SFTs are defined using a background CFT, there will be a \textit{distinct} SFT for every possible set of background fields. 
    \item Within the models of a particular SFT, the different backgrounds are given by solutions to (\ref{EOM}).
\end{enumerate}
The relation between these two notions will be critical in interpreting background independence proofs, and we will discuss this relation in more detail in \S\ref{Sec BI} and \S\ref{Section: Invariant Structure}.

However, to get a simple intuitive picture, we can imagine the spacetime as consisting of a manifold $M$ and some background field configuration $\hat B$.
SFT has fields of all spins, so there are background tensor and spinor fields of all ranks. We will denote such a configuration with the following:
\begin{equation}
    \langle M, \hat B\rangle := \langle M, \hat T, \hat A_{\mu}, \hat g_{\mu\nu}, \dots \rangle.
\end{equation}
The kinematical possibilities of a string field theory are $\langle M, \hat B, \Psi\rangle$. The string field $\Psi $ represents fluctuations around $\hat B$:
\begin{equation}
    \langle M, \hat B, \Psi \rangle := \langle M, \hat T+[\Psi], \hat A_{\mu}+[\Psi]_{\mu}, \hat g_{\mu\nu}+[\Psi ]_{\mu\nu},\dots \rangle,
\end{equation}
where $[\Psi]_{\mu_1\dots \mu_n}$ represent the rank $n$ tensor component of the Fock space expansion of the string field (\ref{eqn: Fock Space Expansion of string field}).\footnote{This is true at linear order, which is sufficient for the sketch we're presenting here, but in general $[\Psi]_{\mu_1,\dots\mu_n}$ will contain non-linear contributions from other components of the string field, and in fact the precise identification depends on the definition of the string vertices $\mathcal{V}$ discussed above; see \citet{mazel2025diffeomorphismclosedstringfield} for more details.}

It is then clear that the solution $\Psi=0$ of the SFT represents $\hat B$, i.e.\ our initial background, in spacetime. Other dynamical possibilities are given by other solutions of the 1PI equation of motion (\ref{EOM}). Consider a specific solution $\Psi'$. The model representing this solution is
\begin{align}
\begin{aligned}
    \langle M, \hat B, \Psi' \rangle &= \langle M, \hat T+[\Psi'], \hat A_{\mu}+[\Psi']_{\mu}, \hat g_{\mu\nu}+[\Psi' ]_{\mu\nu},\dots \rangle \\ &:= \langle M, \hat T', \hat A_\mu ', \hat g_{\mu\nu}',\dots\rangle.
\end{aligned}
\end{align}
However, we could have taken this configuration as our \textit{initial} background configuration; let us denote it as $\hat B'$. We could have formulated a different string field theory around $\hat B'$, say with dynamical variable $\Pi '$. Then the kinematical possibilities of this new theory are
\begin{equation}
     \langle M, \hat B', \Pi' \rangle := \langle M, \hat T'+[\Pi'], \hat A'_{\mu}+[\Pi']_{\mu}, \hat g'_{\mu\nu}+[\Pi' ]_{\mu\nu},\dots \rangle.
\end{equation}
The background $\hat B'$ is represented by $\Pi '=0$ in this new string field theory. It therefore seems that the model $\langle M, \hat B, \Psi'\rangle $ in the first SFT is the same as the model $\langle M, \hat B', 0\rangle $ in the second; both represent the configuration 
\begin{equation}
    \langle M, \hat T', \hat A_\mu ', \hat g_{\mu\nu}',\dots \rangle
\end{equation}
in spacetime. The background independence proofs we will consider in \S\ref{Sec BI} essentially extend this correspondence to all kinematical possibilities of both theories by constructing an isomorphism between them; moreover, dynamical possibilities  will always be taken to dynamical possibilities under such maps. However, these proofs are done at the level of the worldsheet Hilbert space, and the naïve spacetime interpretation presented here likely must be modified, as will be discussed in \S\ref{Section: Invariant Structure}.


\subsection{Witten's cubic theory}\label{SSec: Cubic SFT}

The above closed SFTs are extraordinarily complicated, and in particular, only perturbative solutions to the equations of motion are currently known. Fortunately, for open bosonic strings, there is a much simpler theory given by Witten's cubic string field theory (CSFT) \citep{WITTEN1986253}. The relative simplicity of the theory means that certain \textit{non-perturbative} results are known, which will play some role in questioning the naïve spacetime interpretation above. 

Open strings end on objects called D$p$-branes (henceforth just called D-branes), which live on $p+1$ dimensional submanifolds of the spacetime. The excitations of the open string are then interpreted as fields living on the brane and represent deformations of the brane. From the worldsheet point of view, open strings are Riemann surfaces with boundary. D-branes are then interpreted as different boundary conditions of the worldsheet theory. 

To formulate CSFT, we need an open \textit{and} closed string background, meaning we need a spacetime with a set of background fields and a configuration of D-branes on this spacetime with background fields localized on the D-branes. On the worldsheet, this means we need to choose a boundary action in addition to our CFT, which makes it a boundary conformal field theory (BCFT). Note that the boundary action is constructed using operators from the CFT.

Quantum mechanically, open strings can combine into closed strings, meaning a truncation of string theory to only open strings would be inconsistent in the full quantum string theory. For this reason, we will only work at the classical level (in the sense that we only work at tree level), and there will be no closed string fluctuations at all. In particular, the spacetime itself cannot change in this theory (there is no gravity), and it should instead be seen as a theory describing the worldvolume excitations of D-branes. It is more like a stringy generalization of Yang--Mills theory than general relativity.  



The cubic theory is typically written using an associative $*$-product and takes the form\footnote{For the definition of this associative $*$-product, see \citet{WITTEN1986253}.}
\begin{equation}
    S=\text{Tr}\left( \Psi * Q_B \Psi + \frac 2 3 \Psi * \Psi * \Psi\right),
\end{equation}
which looks like a Chern--Simons action. It can be rewritten in the language of conformal field theory correlation functions,\footnote{See \citet[eq.\ 3.169]{Erler_2022} for the explicit form of this vertex.} where it takes a form more similar to the closed string field theory:
\begin{equation}
    S=\frac 1 2 \bra{\Psi} Q_B \ket{\Psi} +\frac 1 3 \langle \Psi, \Psi,  \Psi \rangle.
\end{equation}
This cubic vertex is very special, as it can cover the entire moduli space of all Riemann surfaces (with boundary) by itself! Thus, this theory can reproduce all open string amplitudes, and we do not need any additional vertices. The equations of motion are 
\begin{equation}\label{eqn: CSFT EOM}
    Q_B \Psi + \Psi * \Psi = 0,
\end{equation}
and the gauge symmetry is 
\begin{equation} \label{eq: cubic gauge symmetry}
    \Psi \to \Psi + Q_B \Psi +  \Psi * \Lambda - \Lambda * \Psi.
\end{equation}

Solutions to the equations of motion correspond to different backgrounds and are interpreted as different D-brane configurations on the spacetime. The simple solution $\Psi=0$ is just our starting configuration. As discussed above, we can construct a different CSFT for every background configuration of D-branes.\footnote{The presentation here is somewhat anachronistic, since the D-brane interpretation of the solutions of the equations of motion \eqref{eqn: CSFT EOM} was available only after the discovery of D-branes, with \citet{Polchinski1989, Horava1989}.}


\section{Background independence}\label{Sec BI}

With this basic background regarding SFT in hand, we turn now to the claims which are made in the literature regarding its background independence. But we do so obliquely, by first considering a simpler, motivating example, to do with the spin-2 reformulation of general relativity. After considering this case in \S\ref{SSec spin-2Theory}, we present a schema for the equivalence of SFTs and their background independence in \S\ref{Ssec: schema for BI}, before turning in \S\ref{SSec: EM solutions} and \S\ref{SSec: Quantum closed string} to two different ways of filling out this schema.

\subsection{Motivating example: spin-2 gravity}\label{SSec spin-2Theory}
Before we tackle claims regarding the background independence of SFT, it will be instructive to review how such ideas play out in a simpler and more familiar setting: that of a spin-2 reformulation of general relativity.\footnote{The example considered in this section is sketched in Erbin's handwritten notes on BI of SFT \citep{Erbin_BINotes}. In this philosophical literature, spin-2 gravity has also been discussed recently by \citet{Linnemann2023, Salimkhani2020-SALTDA-4}; the verions there are not the same as the spin-2 approach considered here.} For simplicitly, we'll consider the case of vacuum general relativity.
Kinematically possible models of vacuum general relativity are given by tuples $\langle M,g_{\mu\nu}\rangle$, where $M$ is a four-dimensional smooth manifold and $g_{\mu\nu}$ a Lorentzian metric. The dynamically possible models are then picked out by the vacuum Einstein field equations $G_{\mu\nu}:=R_{\mu\nu}-\frac{1}{2}Rg_{\mu\nu}=0$, which can be obtained by varying the Einstein--Hilbert action
\begin{equation}\label{eqn EH-action}
    S_{\text{EH}}[g_{\mu\nu}]=\int_M d^4x\ \mathcal{L}_{\text{EH}}[g_{\mu\nu}]=\int_M d^4x\ \sqrt{-g}R[g_{\mu\nu}],
\end{equation}
where $R[g_{\mu\nu}]$ is the Ricci scalar associated to $g_{\mu\nu}$ and $g=\det \left( g_{\mu\nu} \right)$.

Pick now an arbitrary dynamically possible model $\langle M,\hat {g}_{\mu\nu}\rangle$ where $\hat {g}_{\mu\nu}$ is a solution of the Einstein field equations (meaning that $G_{\mu\nu}[\hat{g}_{\rho \sigma}]=0$) and split the metric $g_{\mu\nu}$ in (\ref{eqn EH-action}) as $g_{\mu\nu}=\hat{g}_{\mu\nu}+h_{\mu\nu}$, where $h_{\mu\nu}$ is now an arbitrary two-index tensor and is not in general a metric.
Remember that $\hat{g}_{\mu\nu}$ is a fixed, particular solution to the Einstein field equations meaning that we treat it as a non-dynamical variable and \textit{don't} (at least initially) subject it to variation via Hamilton's principle. Consequently, $h_{\mu\nu}$ subsumes all the interesting dynamics and is the new dynamical variable of $S_{\text{EH}}[\hat{g}_{\mu\nu}+h_{\mu\nu}]$ effectively replacing $g_{\mu\nu}$ in this role. Expanding $S_{\text{EH}}$ as a power series around $\hat{g}_{\mu\nu}$ in terms of $h_{\mu\nu}$ we obtain
\begin{align}\label{eqn EH-expansion}
\begin{aligned}
    S_{\text{EH}}[g_{\mu\nu}] &= S_{\text{EH}}[\hat{g}_{\mu\nu}+h_{\mu\nu}] = S_0[\hat{g}_{\mu\nu},h_{\mu\nu}]+S_1[\hat{g}_{\mu\nu},h_{\mu\nu}]+\dots \\
    &= \int_Md^4x\ \mathcal{L}_{\text{EH}}[\hat{g}_{\mu\nu}] + \int_M d^4x\ \frac{\partial\mathcal{L}_{\text{EH}}[g_{\mu\nu}]}{\partial g_{\mu\nu}}\bigg|_{g_{\mu\nu}=\hat{g}_{\mu\nu}}h^{\mu\nu}+\dots \\
    &= \int_M d^4x\ \sqrt{-\hat{g}}R[\hat{g}_{\mu\nu}] + \int_M d^4x\ \sqrt{-\hat{g}}G_{\mu\nu}[\hat{g}_{\mu\nu}]h^{\mu\nu} + \dots \\ 
    &=: S_{\text{spin-2}}[\hat{g}_{\mu\nu},h_{\mu\nu}].
\end{aligned}
\end{align}
Note that if $\hat{g}_{\mu\nu}$ is a solution to the Einstein field equations as we assumed, then $S_1[\hat{g}_{\mu\nu},h_{\mu\nu}]$ vanishes and we are left only with higher-order terms and $S_0$.

Imagine now an alternative course of history in which the discoveries of early-twentieth century physics unfolded in quite a different fashion.\footnote{In this passage we of course pay homage to our heroes, \citet{Brown2005-RBRPRS-2} and \citet{Stachel}.}
In this alternative course of events, general relativity wasn't conceived of by Albert Einstein and David Hilbert around 1915, but rather by two very different scientists: Albert Beinstein and David Dilbert. Moreover, the formulation of the theory itself happened to be somewhat different from the Einstein--Hilbert formulation. While the \textit{actual} formulation of general relativity---the one described by the Einstein--Hilbert action $S_{\text{EH}}$---is very elegant in the sense that $S_{\text{EH}}$ doesn't require a choice of a fixed background field to be written down, this is not so in the Beinstein--Dilbert formulation (spin-2 theory for short).
In this spin-2 theory, the action is given by $S_{\text{spin-2}}$ and as such requires a choice of a fixed, privileged background $\hat{g}_{\mu\nu}$ in order to be written down. In other words, the Beinstein--Dilbert formulation given by $S_{\text{spin-2}}$ appears to be \textit{background dependent} in a loose, to-be-made-precise sense.\footnote{We set aside the interesting albeit tangential question of why and how Beinstein and Dilbert arrived at the spin-2 theory action given its reliance on a suspect background field, infinitely many terms and overall cumbersome form;
after all, questions of this kind are more properly suited for the historians!} That said, we note one last interesting property of the relationship between this spin-2 theory and  general relativity: the two theories are in an important sense equivalent, because one seems to be just a reformulation of the other since the Einstein--Hilbert action is related to the Beinstein--Dilbert action by a simple change of variables as demonstrated in (\ref{eqn EH-expansion}).

With some ingenuity, Beinstein and Dilbert could have noticed that their action is invariant under a special kind of transformation which maps
\begin{align}\label{eqn EH-shifts}
\begin{aligned}
    \hat{g}_{\mu\nu}&\mapsto\hat{g}'_{\mu\nu}, \\
    h_{\mu\nu}&\mapsto h'_{\mu\nu}:= h_{\mu\nu} +\hat g_{\mu\nu}- \hat{g}'_{\mu\nu}.
\end{aligned}
\end{align}
Here $\hat{g}'_{\mu\nu}$ is another fixed solution to the Einstein field equations. To Beinstein and Dilbert, this invariance would have presumably seemed like a miraculous conspiracy. Each of the infinitely many terms in $S_{\text{spin-2}}$ has very non-trivial dependence on $\hat{g}_{\mu\nu}$ which hides inside the power series coefficients and each successive term in the power series contains increasingly higher powers of $h_{\mu\nu}$ rendering the symmetry transformations (\ref{eqn EH-shifts})  hidden on first blush.\footnote{For recent philosophical work on hidden symmetries, see \citet{BielinskaJacobs, Read2025}.} However, to we followers of Einstein and Hilbert, this invariance of $S_{\text{spin-2}}$ is revealed to be nothing but a manifestation of a trivial property of the Einstein--Hilbert action. For from \eqref{eqn EH-expansion}, it follows immediately that
\begin{align}\label{eqn EH-invariance}
\begin{aligned}
    S_{\text{spin-2}}[\hat{g}'_{\mu\nu},h'_{\mu\nu}] &= S_{\text{EH}}[\hat{g}'_{\mu\nu}+h_{\mu\nu}+\hat{g}_{\mu\nu} - \hat{g}'_{\mu\nu}] \\ 
    &= S_{\text{EH}}[\hat{g}_{\mu\nu}+h_{\mu\nu}]=S_{\text{EH}}[g_{\mu\nu}] = S_{\text{spin-2}}[\hat{g}_{\mu\nu},h_{\mu\nu}],
\end{aligned}
\end{align}
which is the statement of invariance of $S_{\text{spin-2}}$ under (\ref{eqn EH-shifts}). Invariance under (\ref{eqn EH-shifts}) is thus hidden deep down in the intricate structure of $S_{\text{spin-2}}$ and but becomes manifest in $S_{\text{EH}}$.\footnote{This is somewhat akin to Trautman gauge symmetry being a `hidden' symmetry of Newtonian gravitation theory, made manifest in Newton--Cartan theory. See \citet{Read2025}.} Moreover, the fact that such invariance exists allows us to conclude that the choice of $\hat{g}_{\mu\nu}$ in $S_{\text{spin-2}}$ is in some sense irrelevant provided that it is compensated for by appropriate change of the dynamical variables $h_{\mu\nu}$.

We can more formally express this `change of variables' between two different spin-2 theories as follows. Suppose we have two different theories of spin-2 gravity with two different `background metrics' $\hat g_{\mu\nu}$ and $\hat g'_{\mu\nu}$. The dynamical variable in the first theory will be denoted $h_{\mu\nu}$, whereas the dynamical variable in the second theory will be $h'_{\mu\nu}$. Let us define (suppressing indices)
\begin{align}
\begin{aligned}
    S_1[h] &:= S_{\text{spin-2}}[\hat g, h], \\
    S_2[h'] &:= S_{\text{spin-2}}[\hat g', h'].
\end{aligned}
\end{align}
Both $S_1$ and $S_2$ are defined on the space of sections of symmetric two-index tensors on the manifold: $S_i:\Gamma(\text{Sym}(T^*M\otimes T^*M))\rightarrow \mathbb{R} $. Now, let a bijection $F:\Gamma(\text{Sym}(T^*M\otimes T^*M))\rightarrow\Gamma(\text{Sym}(T^*M\otimes T^*M)) $ (the `change of variables map') be defined by $F(h)=h-\hat g+\hat g'$, i.e.\ \eqref{eqn EH-shifts}; then, by the above argument, the pullback of $S_2$ under this bijection is $S_1$:
\begin{equation} \label{COV}
   F^*S_2[h]:= S_2[F(h)]=S_1[h].
\end{equation}
Again, this is simply a formal way to say that the two actions are related by a change of variables.

The kinematically possible models of the first theory with action $S_1$ are $\langle M, {\hat g}_{\mu\nu}, h_{\mu\nu} \rangle$, while those of the latter theory with action $S_2$ are $\langle M, \hat{g}_{\mu\nu}', h_{\mu\nu}' \rangle$. These are related to the corresponding model $\langle M, g \rangle$ in general relativity by 
\begin{equation}\label{eqn: graviton}
    g_{\mu\nu}= \hat g_{\mu\nu}+h_{\mu\nu} = \hat g'_{\mu\nu} + h'_{\mu\nu}.
\end{equation}
We then have the following action-preserving isomorphism of models between the two spin-2 theories:
\begin{equation}
    \langle M, \hat g, h \rangle \sim \langle M, \hat g', h + \hat g - \hat g' \rangle.
\end{equation}
Since the action is preserved under this map, the dynamically possible models are taken to dynamically possible models in this isomorphism. If this isomorphism really represents a physical equivalence between models of different theories, then presumably only some invariant structure between these models should be taken to be physical. In the case of spin-2 gravity, this is of course given by the metric of general relativity \eqref{eqn: graviton}
(modulo diffeomorphisms).  

We'll soon see that there are many analogies between the situation in SFT and our spin-2 theory toy example. In fact, Sen and Zwiebach's proofs of background independence of SFT---which we take up and analyze in \S\ref{SSec: Quantum closed string}---can be seen as analogous to the ingenious step of seeing the non-trivial invariance of $S_{\text{spin-2}}$ under (\ref{eqn EH-shifts}) \textit{without} the prior knowledge of $S_{\text{EH}}$ and its relationship to $S_{\text{spin-2}}$ encapsulated by (\ref{eqn EH-expansion}).

\subsection{Schema for equivalence of SFTs and BI proofs}\label{Ssec: schema for BI}

Background independence proofs in SFT turn out to have many parallels to the spin-2 case presented above. In particular, the overarching goal is to show that two different SFTs formulated around different backgrounds are actually equivalent. In this section, we will describe a schema for showing when two string field theories are equivalent. This schema will hold for all string field theories, closed or open.

In parallel with the spin-2 case, for two SFTs to be equivalent, we require some kind of isomorphism of models between the two theories, 
\begin{equation}
    \langle M_1, \hat B_1, \Psi _1\rangle \sim \langle M_2, \hat B_2, \Psi_2\rangle.
\end{equation}
Here, $M_i$ is the spacetime manifold, $\hat B_i$ is the background structure of the SFT, and $\Psi_i$ is the dynamical string field. It is then argued that since these models are physically equivalent, the true physical structure must be invariant between these models. Since $\hat B_1$ and $\hat B_2$ are different in these models, it is argued that they are not physical, i.e.\ the background choice is not physical. Hence, everything physical in the string field theory is independent of the choice of background.

We don't just want an arbitrary isomorphism of models, however; this of course can be achieved for any two theories whose solution spaces are of the same dimension/cardinality. Some salient dynamical structure must be preserved under the mapping. In the classical case, by analogy with spin-2 gravity, we  require that the actions be equivalent under a pullback, i.e.\ that the different actions be related by a change of variables. In addition, due to the gauge symmetry of SFT, we also need it to be the case that the BV structures of the two theories match. In the quantum case, we want the map to preserve all correlation functions, the path integral, and the BV structure of the quantum master action. Equivalently, we can preserve the 1PI equations of motion. This will ensure that dynamical possibilities are taken to dynamical possibilities and that the models are empirically equivalent.

(Before continuing, it is worth making here two philosophical asides. First, the idea of finding a mapping between the models of two SFTs such that dynamics and empirical content are preserved means that these approaches to the background independence of SFT can be understood through the lens of the recent philosophical literature on dualities---see \citet{DHButterfield} for a book-length exposition, and \citet{DeHaro2019-DEHTEA} for an attempt to cash out a novel notion of theoretical equivalence in terms of the `schema for dualities' presented in that former work. The point here is that these mappings between SFTs exactly fit the mould of dualities in this sense---although note that it is not typical in the physics literature to describe these SFTs as `dual'. Our second point is this: a philosopher might hanker after more details regarding what `empirical equivalence' means in this context. To be clear: in this article, we always mean `empirical equivalence' in the standard philosophers' sense of `same empirical content', which is a strictly weaker notion than `physical equivalence', which means `describe the same possible world or subpart thereof'.) 

One of the key conceptual ideas of the background independence proofs of \citet{ZweibachSen, ZweibachSen2} is that, since the two models $\langle M_1, \hat B_1, \Psi _1\rangle$ and $\langle M_2, \hat B_2, \Psi_2\rangle$ will be empirically equivalent (since the actions or correlation functions will be the same), they should \emph{eo ipso} be regarded as being \textit{physically} equivalent.\footnote{Readers familiar with the literature on the philosophy of symmetries will recognise this as something like `interpretationalism' about symmetries, according to which one can regard empirically equivalent, symmetry-related models of physical theories \emph{ab initio} as being physically equivalent. See \citet{TMN, JRTMN} for details.} Then, the backgrounds $\hat B_1 $ and $\hat B_2$ have no physical significance; presumably, only some invariant structure between these models will be physical, just as with the dynamical metric $g_{\mu\nu}$ in the case of spin-2 gravity and general relativity. We will discuss what this invariant structure could be in \S \ref{Section: Invariant Structure} below. Note that this is a sort of generalization of the debates about symmetry: there, symmetry-related models \textit{within} some theory are identified, whereas in the SFT case, we wish to identify models \textit{between} two (or more) theories.\footnote{This is the kind of case considered by \citet{JRTMN} in the case of dualities in physics.} In this sense, SFT equivalences are something like `inter-theoretic' gauge equivalences; the different backgrounds and corresponding actions are different `gauge choices' which are gauge equivalent, albeit not manifestly so.\footnote{To pick up on our above point regarding dualities: the connection between dual theories and gauge redundancy has been elaborated by \citet{Rickles2017-RICDTS}; what we say here is consistent with the lessons presented in that article.} 

With all of this in mind, let's now give the general schema for the equivalence of SFTs. Recall that a given SFT, call it SFT$_i$, is specified by the following data:
\begin{enumerate}
    \item The worldsheet CFT, which comes equipped with a  Hilbert space $\mathcal{H}_i$ and Fock space basis $\ket{\phi_i}$.\footnote{We assume $\mathcal{H}_i $ is the part of the more general CFT Hilbert space which satisfies the level matching conditions.}
    \item A dynamical string field $\Psi_i \in \mathcal{H}_i$.
    \item The SFT action $S_i: \mathcal{H}_i \to \mathbb{R}$.
\end{enumerate}

Let us now have two string field theories SFT$_1$ and SFT$_2$ with an isomorphism (up to gauge transformation\footnote{The isomorphism takes gauge equivalence classes to gauge equivalence classes---recall \eqref{eqn: closed superstring gauge invariance}. Actually, in general, we only need an isomorphism between cohomologies (gauge equivalence classes) of the theories, and it might not be possible to have a direct isomorphism of Hilbert spaces. This is the case for the Erler--Maccaferri solutions discussed in the next subsection. One can also phrase this requirement as saying that the BV form is preserved, $\omega_1 = f^* \omega_2$.}) between their Hilbert spaces:
\begin{align} \label{Isomorphism}
\begin{aligned}
    f&: \mathcal{H}_1 \to \mathcal{H}_2 \nonumber \\
    f^{-1}&: \mathcal{H}_2 \to \mathcal{H}_1.
\end{aligned}
\end{align}
As already mentioned above, the kinematically possible models of a string field theory are determined by (the gauge equivalence classes of) $\Psi_i$, which is an arbitrary member of $\mathcal{H}_i$. More precisely, the models are given by $\langle M_i, \hat B_i, \Psi_i\rangle$, where $\hat B_i$ is the background structure determined by the underlying worldsheet CFT; since this is fixed, every distinct model is in one-to-one correspondence with the (gauge equivalence classes of the) members of the Hilbert space. Thus, if we have such an isomorphism of Hilbert spaces, then we already have an isomorphism of models. In order for the theories to be equivalent, however, this isomorphism must preserve the action under a pullback, meaning the two theories are related by a change of variables:
\begin{equation} \label{BI Condition}
    f^*S_2[\Psi_1] \equiv S_2[f(\Psi_1)] = S_1[\Psi_1].
\end{equation}
For two SFTs to be equivalent, we just need there to be an isomorphism up to gauge transformations $f: \mathcal{H}_1 \to \mathcal{H}_2$ that satisfies \eqref{BI Condition}. This will ensure that every model in SFT$_1$ has an empirically(/physically) equivalent counterpart in SFT$_2$ and vice versa.

This map $f$ will typically consist of three parts (all isomorphisms up to gauge transformations):
\begin{enumerate}
    \item A translation $T:\mathcal{H}_1\to \mathcal{H}_1$.
    \item A field redefinition $G: \mathcal{H}_1\to \mathcal{H}_1$.
    \item An abstract Hilbert space isomorphism $F: \mathcal{H}_1\to\mathcal{H}_2 $.
\end{enumerate}
Then, $f=F\circ G\circ T: \mathcal{H}_1\to\mathcal{H}_2$ will be the action-preserving isomorphism in (\ref{BI Condition}). 

The translation $T$ can be thought of as shifting to the solution in SFT$_1$ that represents the background configuration $\hat B_2$ of SFT$_2$. To find it, one must solve the 1PI equation of motion (\ref{EOM}). Let $\hat \Psi_2\in \mathcal{H}_1$ be this solution; then we expand around it (just as in the spin-2 case, where we expand around a particular background metric): $\Psi_1 = \hat \Psi_2+\Phi_1$. This $\Phi_1$ will be our new dynamical variable, and $T$ implements this shift:
\begin{equation}
    T(\Psi _1)=\Psi_1-\hat \Psi_2=\Phi_1.
\end{equation}
The field redefinition $G$ ensures that the form of the actions is the same for both theories, and the map $F$ will be a general isomorphism between the two Hilbert spaces of the CFTs. In some cases, it can be interpreted as a `connection on the space of CFTs' (see \citet{ZweibachSen}).





In the quantum case, instead of equating the actions, we equate the whole path integral with measure (\cite{ZweibachSen}):
\begin{equation}\label{eqn: Path Integral BI}
    \mathcal{D}\Psi_1 e^{S_1[\Psi_1]} = f^*(\mathcal{D}\Psi_2 e^{S_2[\Psi_2]}). 
\end{equation}
This will ensure the equivalence of all correlation functions as long as the gauge equivalence classes are preserved, i.e. if the BV form is preserved. To see this, let $A_2$ be an operator in $\mathcal{H}_2$ which satisfies the BV condition. Then $A_1:=f^*A_2$ will also satisfy the BV condition in $\mathcal{H}_1$ and will therefore be a physical observable. Moreover,
\begin{align}
\begin{aligned}
     \langle A_1\rangle &= \int_{L_1} \mathcal{D}\Psi_1 A_1(\Psi_1)e^{S_1}\\ &=\int_{f^*L_2} f^*\left( \mathcal{D}\Psi_2 A_2(\Psi_2)e^{S_2} \right) \\&=\int_{L_2}  \mathcal{D}\Psi_2 A_2(\Psi_2)e^{S_2}  =\langle A_2\rangle.
\end{aligned}
\end{align}
To go from the second to the third line, we note that the pullback simply expresses a change of variables in the path integral. Thus, we find that the correlation functions of the two theories are equivalent. Note that equating the path integral (\ref{eqn: Path Integral BI}) is identical to equating the 1PI action under a pullback. This can be seen from the definition of the 1PI action (\ref{eqn: 1PI}), which involves performing the full path integral over the classical fields through the dependence on $Z[J]$. Thus, the quantum case works identically to the classical case with the simple replacement of the action with the 1PI action.

Instead of requiring (\ref{eqn: Path Integral BI}), we can follow \citet{Sen2018} in equating the 1PI equations of motion under the following:
\begin{align}
      Q_{B,1} \ket{\Psi_1}+\sum_{n= 1}^\infty \frac 1 {n!} \mathcal{G}_1 \ket{[\Psi_1^n]}&=f^{-1}\left( Q_{B,2} \ket{f(\Psi_1)}+\sum_{n= 1}^\infty \frac 1 {n!} \mathcal{G}_2 \ket{[f(\Psi_1)^n]}\right).
\end{align}
This will also lead to an empirical equivalence of the two theories, and dynamical possibilities will be taken to dynamical possibilities, since the equations of motion are preserved (which are what classify the dynamical possibilities). Sen's original proof of background independence of closed superstring field theory needed to follow this strategy of equating the equations of motion rather than the actions. The reason is due essentially to the doubled string field $\tilde \Psi$; indeed, one can prove that Sen's action (\ref{eqn: superstring field theory action}) \textit{cannot} be background independent in the sense of equivalence of (1PI) actions, see \citet{Sen2018}. However, recently, a new action for superstring field theory has been found by \citet{hull2025newactionsuperstringfield}. With this new formulation, it is possible to directly equate the 1PI actions, and it is no longer necessary to invoke the equations of motion.  

Of course, the above is but a sketch, and in practice the challenge lies in identifying/constructing the isomorphism $f=F\circ G\circ T: \mathcal{H}_1\to\mathcal{H}_2$. Over the next two subsections, we'll look at implementations of this schema to the open string (\S\ref{SSec: EM solutions}) and closed string (\S\ref{SSec: Quantum closed string}). In each case, to repeat, what's involved is explicitly constructing this $f$.

\subsection{The Erler--Maccaferri solutions}\label{SSec: EM solutions}


Let us now implement this schema for the open string. Throughout this section, we will follow the discussion of \citet{ErlerMaccaferri1, ErlerMaccaferri2}. Recall from \S\ref{SSec: Cubic SFT} Witten's (classical) cubic action for a CSFT,
\begin{equation}
    S_1=\text{Tr}_1\left(\Psi_1 *Q_{B1}\Psi_1 +\frac 2 3 \Psi_1 *\Psi_1 *\Psi_1\right).
\end{equation}
This action requires background closed \textit{and} open string fields; we will denote them as $\hat C_1$ and $\hat O_1$ respectively. Consider another CSFT, say with action $S_2$ and dynamical variable $\Psi_2$, formulated around a different open string background, but with the same closed string background: $\hat C_1=\hat C_2$, $\hat O_1\neq \hat O_2$. In other words, consider a CSFT formulated around a different D-brane configuration in the same spacetime. 

The condition for background independence (\ref{BI Condition}) requires that there exist a map $f$ between the Hilbert spaces of the two CSFTs which preserves the action. As we have seen, a typical map $f$ will involve a translation (finding the solution) $T$, a field redefinition $G$, and a map $F$ between the Hilbert spaces. For the case of CSFT, the difficult part of proving the existence of $f$ is in finding the solutions corresponding to different backgrounds, i.e.\ finding $T$. However, \citet{ErlerMaccaferri1, ErlerMaccaferri2} explicitly constructed all \textit{non-perturbative} solutions to the CSFT equations of motion which represent other CSFT backgrounds. There is also a natural map $F$ between the Hilbert spaces of all pairs of CSFTs with the same $\hat C$, which is given by open strings connecting the D-branes. The field redefinition $G$ is simply the identity in this case. 

If one expands in some CSFT$_1$ around the Erler--Maccaferri solution $\hat \Psi _2^{\text{EM}}$ representing the other CSFT$_2$ background $\hat O_2$, then uses the map $F$ (denoted $F(\Phi)=\bar \Sigma \Phi\Sigma $ in this case), one can show that the actions $S_1$ and $S_2$ are equivalent under the pullback of $f=F\circ T$: $f^*S_2[\Psi_1]=S_1[\Psi_1]$; moreover, this map preserves gauge equivalence classes under (\ref{eq: cubic gauge symmetry}). Thus, we have an action-preserving isomorphism between the models of any two CSFTs formulated with the same closed string background:
\begin{equation}
    \langle M, \hat C; \hat O_1, \Psi_1\rangle \sim \langle M, \hat C;\hat O_2, \Psi_2\rangle. 
\end{equation}
If we accept this criterion as amounting to physical equivalence, then it seems that the choice of open string background $\hat O_i$ is irrelevant to the physics of CSFT and hence should be given no ontological significance. The theory is still clearly background dependent because of $\hat C$, but this argument suggests that some of the other structure needed in formulating the theory is superfluous. In the open string case, it is possible to give a formulation which is manifestly independent of $\hat O$, as will be discussed in \S\ref{SSec: Witten BSFT}.

\subsection{The quantum closed string}\label{SSec: Quantum closed string}
We will now use the schema for both the bosonic and supersymmetric closed string field theories, following closely \citet{ZweibachSen, ZweibachSen2} and \citet{Sen2018}. These are arguably most interesting from the perspective of background independence, since they are quantum theories where spacetime will also fluctuate. However, due to the complexity of the theories, we lack many of the results present for the open string. In particular, there is no known analogue of the Erler--Maccaferri solutions, as it is simply not known how to solve analytically the closed string field theory equations of motion for backgrounds `far away' from our starting configuration. However, there is still an isomorphism of models for `infinitesimally close' backgrounds. From the CFT point of view, these backgrounds are related by small marginal deformations. From the spacetime point of view, the backgrounds $\hat B$ and $\hat B'$ differ by infinitesimally small field values. 

\citet{ZweibachSen, ZweibachSen2} and \citet{Sen2018} prove the background independence of bosonic closed string field theory and closed superstring field theory respectively for such `infinitesimal' shifts in backgrounds. The map $T$ can be found by recursively solving the equations of motion (\ref{EOM}) order-by-order in some small parameter $\epsilon$. Then, the map $F$ is given by a \textit{connection on the space of CFTs}: one can view CFTs related by small marginal deformations as living on a manifold with the coordinates representing coupling constants in front of the various possible marginal deformations. The CFT Hilbert spaces form a bundle over this space, and the connection $F$ is an isomorphism between the Hilbert spaces at different points. The most technically challenging part of the proof is the existence of a $G$ that transforms the 1PI action (or equations of motion) into each other.\footnote{As mentioned above, Sen originally needed to equate equations of motion rather than actions due to the auxilliary field $\tilde \Psi$. Due to \citet{hull2025newactionsuperstringfield} this is no longer necessary, and one can simply equate the 1PI actions.} This is done with an explicit construction that is also a recursive series in orders of $\epsilon$. One must show that there are no topological obstructions to this procedure. Once this is established, the existence of a map $f$ between the Hilbert spaces that preserves the action is shown, and hence there is an action-preserving isomorphism of models
\begin{equation}
    \langle M, \hat B_1, \Psi _1\rangle \sim \langle M, \hat B_2,  \Psi _2\rangle
\end{equation}
for infinitesimally close backgrounds $\hat B_1$ and $\hat B_2$. 

\citet{ZweibachSen, ZweibachSen2} also give an argument that one can `integrate' such infinitesimal shifts to get finite shifts, and hence that the above isomorphism should extend to closed SFTs formulated around backgrounds $\hat B$ and $\hat B'$ that are a finite distance apart. This amounts to solving a certain differential equation in the deformation parameter, and Sen and Zweibach prove that there is no topological obstruction in solving this equation. However, there may potentially be divergences in this process which are interpreted as indicating the breakdown of the variables of SFT$_1$ in describing the background SFT$_2$, as discussed by \citet{Sen2018}.

\section{Invariant structure}\label{Section: Invariant Structure}

The background independence proofs from the previous section suggest that some of the structure used in formulating the various string field theories is superfluous. Explicitly, we find that some models are isomorphic: $\langle M_1, \hat B_1, \Psi _1\rangle \sim \langle M_2, \hat B_2, \Psi _2\rangle$. There is therefore a case to be made---or at least explored---that the physical content of the string field theory is whatever is invariant under these isomorphisms. In this section, we will explore various candidates for what amounts to such `background independent' structure of SFT. 

In \S\ref{SSec invariant sugra}, we analyze the most straightforward candidate for a background independent structure in closed string field theory. In \S\ref{SS invariant YM}, we repeat a similar analysis for the open string and discuss some issues which arise. In \S\ref{SSec: Witten BSFT} we review a different formulation of open string field theory which resolves these issues and analyze the invariant content in \S\ref{BSFT analysis}. We then review an analogous formulation of closed string field theory in \S\ref{SS: cZ review} and analyze its invariant content in \S\ref{SS: cZ analysis}. Finally, we summarize the results in \S\ref{SS: models of SFT}.

\subsection{Invariant (super)gravity structure}\label{SSec invariant sugra}

We now analyze the closed string, both classical and quantum.\footnote{The only difference in the analyses is whether we use the classical equation of motion or the 1PI equation of motion to identify the DPMs. The latter has $g_s$ corrections taken into account.} As discussed in \S\ref{SSec: Quantum closed string}, we have an isomorphism of models between SFTs formulated around different backgrounds $\hat B_1$ and $\hat B_2$,
\begin{equation}\label{eq:iso}
    \langle M, \hat B_1, \Psi _1\rangle \sim \langle M, \hat B_2,  \Psi _2\rangle.
\end{equation}
To understand what this means in more detail, let us consider the spacetime interpretation of a general state in the Hilbert space of one of the string field theories, $\Psi _1 \in \mathcal{H}_1$. As discussed in \S\ref{SSec: Backgrounds in SFT}, this will correspond to some set of fields in spacetime, which we can view as a perturbation around the initial background $\hat B_1$; in the notation of \S\ref{SSec: Backgrounds in SFT}, the spacetime model is 
\begin{equation}
\langle M, \hat B_1, \Psi _1 \rangle = \langle M, \hat B_1 +\Psi_1\rangle:= \langle M, \hat T_1(x) + [\hat \Psi_1(x)], \hat g_{1{\mu\nu}}(x)+[\hat \Psi_1(x)]_{\mu\nu}, \dots \rangle. 
\end{equation}
The isomorphism of models between the string field theories \eqref{eq:iso} tells us that for every spacetime configuration of fields in SFT$_1$, there is a corresponding empirically equivalent model in SFT$_2$. This empirically equivalent model will have a similar spacetime representation as 
\begin{equation}
    \langle M, \hat B_2 + \Psi _2\rangle =\langle M, \hat T_2(x)+[\Psi_2(x)], \hat g_{2\mu\nu}(x)+[\Psi_2(x)]_{\mu\nu}, \dots \rangle. 
\end{equation}
For these models to be equivalent, it is natural to equate the fields of each rank up to gauge transformations (which we denote by `$\simeq$'):
\begin{gather}
    \hat T_1 (x) + [\hat \Psi_1(x)] \stackrel{?} {\simeq} \hat T_2 (x) + [\hat \Psi_2(x)] \\ \label{eqn: grav equiv} \hat g_{1{\mu\nu}}(x)+[\hat \Psi_1(x)]_{\mu\nu} \stackrel{?} {\simeq} \hat g_{2\mu\nu}(x)+[\Psi_2(x)]_{\mu\nu} \\ \nonumber \vdots 
\end{gather}
The latter equation (\ref{eqn: grav equiv}) bears a striking resemblance to (\ref{eqn: graviton}) for spin-2 gravity. This similarity suggests that the background independent structure of a model of SFT will be a combination of the background field with the fluctuation, which will simply be the value of the spacetime field $g_{\mu\nu}(x):=\hat g_{1\mu\nu}(x) +[\Psi_1(x)]_{\mu\nu}= \hat g_{2\mu\nu}(x) +[\Psi_2(x)]_{\mu\nu}$.
This will also be the case for the other fields with different ranks, i.e., the background independent model will consist of the sum of the background with the appropriate component of the string field, which will give us the value of the spacetime field. Thus, the background independence proofs suggest that the models of string field theory should be taken to be of the form
\begin{equation}\label{eqn: sugra invariants}
    \langle M, \hat B + \Psi \rangle = \langle M, \hat T+[\Psi], \hat g_{\mu\nu}+[\Psi]_{\mu\nu}, \dots \rangle =\langle M, T(x), g_{\mu\nu}(x), \dots \rangle.
\end{equation}
However, note that the latter models are precisely the same as those of a gravitational theory with a matter sector which comes from the various string modes. In the case of the bosonic string, these will be a 26D dilaton gravity theory with massive fields of higher spins (plus the tachyon), and in the case of the superstring, these will be the models of the appropriate supergravity extended by similar stringy massive fields.

It therefore seems that the invariant content of the kinematically possible models of SFT is really that of (super)gravity with the matter content being given by the various other string excitations. That said, the dynamically possible models are slightly different from classical, minimally coupled (super)gravity due to $\alpha'$ corrections (which apply to both classical and quantum string field theory) and $g_s$ corrections (which only apply to quantum string field theory) to the equations of motion coming from stringy interactions. In this sense, SFT is `just' an extended, UV complete version of spin-2 gravity.
It therefore seems that the invariant content of the kinematically possible models of SFT is really that of (super)gravity with the matter content being given by the various other string excitations.

There are, however, some reasons to call into question this natural, invariant (super)gravity structure. In particular, the background independence proofs relied heavily on the Hilbert spaces of the worldsheet CFTs. Although these Hilbert spaces have Fock space bases which allow straightforward spacetime interpretations of models \textit{within} a specific theory, they can lead to some puzzling results for spacetime interpretations when comparing Fock space bases \textit{between} theories, particularly when comparing backgrounds that are `far away'. However, this is precisely what we are interested in when assessing background independent structure. To illustrate this point, we will construct the analogous `natural structure' for Witten's cubic string field theory and discuss some issues that occur when considering the Erler--Maccaferri solutions. 


\subsection{Invariant Yang--Mills structure}\label{SS invariant YM}


Let us try to apply analogous reasoning from above to the open string field theory. Open string fields living on D-branes are localized to the submanifold of the D-brane. Thus, for example, if we have a D24-brane on a submanifold $\Xi \subset \mathbb{R}^{25,1}$ of 26D Minkowski space, the fields of the brane will be localized to $\Xi$. The CSFT formulated on this brane describes open string fluctuations of the brane. We will use $x$ as a coordinate on all of Minkowski space and $\xi$ as a coordinate on $\Xi$. The background of our CSFT will be some collection of closed and open fields; we will separate them with a semicolon for clarity, with closed background fields on the left and open background fields on the right:  
\begin{equation}
    \langle M, \hat C ; \hat O \rangle = \langle M, \hat T(x), \hat g_{\mu\nu}(x), \dots ; \hat T(\xi ) , \hat A_{\mu}(\xi), \dots \rangle.
\end{equation}
The open string field $\Psi$ will leave the closed string background unchanged, but will describe fluctuations of the fields living on the brane. Thus, a general model of this CSFT will be 
\begin{equation}
    \langle M, \hat C ; \hat O, \Psi \rangle = \langle M, \hat T(x), \hat g_{\mu\nu}(x), \dots ; \hat T(\xi ) +[\Psi(\xi)] , \hat A_{\mu}(\xi)+[\Psi(\xi)]_\mu , \dots \rangle.
\end{equation}
Using the same reasoning as above, the natural background independent structure, or rather, the structure independent of the open string background, should be the sum of the open string background and the fluctuation:
\begin{align}
    \langle M, \hat C ; \hat O +\Psi \rangle& := \langle M, \hat T(X), \hat g_{\mu\nu}(X), \dots ; \hat T(\xi ) +[\Psi] , \hat A_{\mu}(\xi)+[\Psi]_\mu, \dots \rangle \nonumber \\&:=
\langle M, \hat T(X), \hat g_{\mu\nu}(X), \dots ; T(\xi) , A(\xi)_, \dots \rangle.
\end{align}
In other words, the `open string background independent structure' should be the values of the fields living on the brane, $T(\xi), A_\mu (\xi), \dots$. This is also the field content of a 24D Yang--Mills theory, again with the appropriate matter sector coming from higher string modes. 

The problem arises when we compare this model to a different CSFT which is formulated around some other D-brane configuration with the same closed string background. Let us consider the analogous `background independent structure' for a CSFT formulated around some D23-brane which lives on a submanifold $\Sigma \subset \Xi$. The fields of the D23-brane will be localized to $\Sigma$, which has coordinates $\sigma$. Thus, the models are 
\begin{equation}
    \langle M, \hat C ; \hat O' + \Psi' \rangle = \langle M, \hat T(x), \hat g_{\mu\nu}(x), \dots ; T'(\sigma), A'_\mu(\sigma), \dots \rangle. 
\end{equation}
These are the models of a 23D Yang--Mills theory with massive stringy fields. 

This is perhaps not surprising: the effective theory on the worldvolume of a D$p$-brane is simply $p$-dimensional Yang--Mills theory. CSFT is, in a sense, a `UV completion' of this Yang--Mills theory. However, the Erler--Maccaferri solutions make the rather astonishing suggestion that all CSFTs with the same closed string background are equivalent, and hence, all of these various Yang--Mills theories can arise as effective theories from a CSFT of naïvely different dimension, and even different gauge group. For instance, the simplest configuration of the D23-brane with no excitations (representing a vacuum of the 23D Yang--Mills theory/CSFT) arises as a `lump' solution within the 24D Yang--Mills theory/CSFT. Moreover, multi D-brane solutions (which can represent $U(N)$ gauge theories) can arise as solutions of the CSFT of a single brane, which only has $U(1)$ gauge symmetry. Thus, the Erler--Maccaferri solutions seem to suggest that even the spacetime fields are not the correct background independent variables for string field theory.

In the closed string case, analogous solutions that are `far away' are not known, so the spacetime field approach seems more plausible as an account of background independent structure. However, the above argument from the open string suggests that the supergravity models are not sufficient for the full closed SFT, especially when identifying models which may have different matter fields or spacetime topologies. It is therefore important to look at other candidates for background independent structure.

\subsection{A theory on the space of theories: Witten's BSFT}\label{SSec: Witten BSFT}

In this section, we will briefly review a formulation of open string field theory which seems to be manifestly independent of the superfluous background structure from earlier; this formulation is called the \textit{boundary string field theory} (BSFT) (\cite{Witten}). 

String fields take values in the Hilbert space of the worldsheet theory $\mathcal{H}$. Moreover, the background independence proofs from above also rely on isomorphisms between these worldsheet Hilbert spaces. Thus, it is plausible that the background independent structure of SFT could be formulated in terms of worldsheet data rather than spacetime data. Indeed, there is a formulation of open string field theory, which in a precise sense is a `theory on the space of worldsheet theories'. Unfortunately, it is not clear how well-defined this space is due to worries about UV divergences and non-renormalizable boundary interactions. Nonetheless, as we will see, it is possible to write down an action on this space which is manifestly independent of the open string background $\hat O$, though it still depends on a fixed closed string background. 

Recall that open string field theory is a classical theory, which means we ignore string loops. The worldsheet of an open string without loops is conformally equivalent to a disk $D$. Consider a worldsheet theory with an action 
\begin{equation}
    S=\int_{D} d^2\sigma \left (\hat g^{\mu\nu}\partial_\mu X^a \partial _\nu X_a +\dots \right) + \int _{\partial D } d\sigma \mathcal{V}.
\end{equation}
This is specified by some closed string background $\langle \hat g_{\mu\nu}, \dots \rangle$ and a boundary Lagrangian $\mathcal{V}$, which is a local, ghost number zero operator constructed from the `bulk' CFT.\footnote{By the bulk CFT, we mean the 2D CFT on the disk specified by the first term in the worldsheet action, i.e., the closed string background. We are not talking about holographic bulk/boundary theories.} As with the CSFT, we will work with a fixed closed string background, meaning we will fix the first term of the above action. 

Note that in general, the boundary term $\mathcal{V}$ can break conformal invariance. Imposing conformal invariance of the boundary action gives us classical equations for open string background fields, just like how imposing conformal invariance of the `bulk' CFT gives us equations for closed string background fields, as discussed in \S\ref{SSec: Worldsheet backgrounds}. In particular, instead of the Einstein field equations, we get Maxwell's equations (\cite{ABOUELSAOOD1987599}).

Now, notice that the space of possible boundary Lagrangians $\mathcal{V}$ is just the space of ghost number zero operators of the bulk CFT. Thus, in this sense, the ghost number zero operators of the bulk CFT parameterize the space of possible boundary theories for a given closed string background. To see this more explicitly, one can find a basis of this space $V_i$ and expand the general Lagrangian as $\mathcal{V}=\sum_i \lambda_i V_i$; the coefficients $\lambda_i$ are coupling constants, and they are the variables of the BSFT. The goal would be to write down an action on this space so that the `on-shell' $\mathcal{V}$ are the classical open string backgrounds. For technical reasons, however, it is not possible to write an action directly on the space of ghost number zero operators. One must move to the space of ghost number $1$ operators $\mathcal{O}$ such that $\mathcal{V}=b_{-1}\mathcal{O}$, where $b_{-1}$ is defined in terms of the bulk $b$ ghost. This will still give us all possible (matter) boundary actions if we choose $\mathcal{O}=c\mathcal{V}$, but there is some redundancy, as multiple $\mathcal{O}$ can represent the same $\mathcal{V}$.\footnote{For example, this can be done by shifting $\mathcal{O} \rightarrow \mathcal{O}+\Delta$, where $b_{-1}\Delta =0$.}

On the space of ghost number $1$ operators, Witten's action is implicitly defined by
\begin{equation}\label{eq:BSFT-action}
    dS_{\text{BSFT}} = \frac 1 2  \oint \oint d\sigma_1 d\sigma_2\langle d\mathcal{O}(\sigma_1) \{Q_B, \mathcal{O}(\sigma_2)\}\rangle,
\end{equation}
where $Q_B$ is the bulk BRST charge, and $d$ is an exterior derivative on the space of couplings in the boundary Lagrangian. The extremal configurations ($dS=0$) are those that satisfy $\{Q_B, \mathcal O\} = 0$. These in turn are precisely the conformally invariant boundary actions (at least in cases where ghosts and matter decouple). Thus, Witten's action tells us that the classical backgrounds of open string theory are given by conformally invariant boundary theories! It is exactly this requirement of worldsheet conformal invariance which gives us the spacetime equations of motion. Also, notice that this action manifestly does not refer to any open string background $\hat O$, unlike the cubic string field theory from earlier. 

Ultimately then, there's a case to be made that what we have here is a \emph{manifestly} background independent formulation of open SFT; the analogue of the Einstein--Hilbert action for general relativity in the spin-2 gravity case.\footnote{The explicit relation between BSFT and CSFT is explained in \citet{Chiaffrino_2019} and \citet{totsukayoshinaka2025bsftlikeactioncohomomorphism}.} This, of course, goes beyond the invariant structures offered in the previous two subsections.


\subsubsection{Models of BSFT}\label{BSFT analysis}

Let us now analyze the models of BSFT. The kinematic possibilities are spanned by the ghost number 1 operators of the bulk CFT. However, we will represent the models in terms of the boundary Lagrangian $\mathcal{V}$ directly, since the interpretation of the ghost number 1 operators $\mathcal{O}$ remains somewhat obscure, and we can always go from the ghost number operator to a boundary Lagrangian using $\mathcal{V}=b_{-1}\mathcal{O}$.

In the case where the bulk CFT is a non-linear sigma model (`NLSM'), $\mathcal{V}$ will represent some D-brane configuration within the spacetime, perhaps with some non-zero value for worldvolume fields within the brane. We will simply denote the models in this case as $\langle M, \hat C, \mathcal{V} \rangle$, where the operator $\mathcal{V}$ is taken to represent some D-brane configuration within $M$.

\subsection{The $cZ$ action}\label{SS: cZ review}

We will now review a similar `manifestly background independent' formulation of classical closed string field theory. BSFT liberates open string field theory from the requirement of an initial D-brane configuration by living on the space of worldsheet theories; the equation of motion of the BSFT action tells us that on-shell actions are conformally invariant. It is natural to ask whether the same move can be made for the closed string. That is, can we write down an action on the space of closed string worldsheet theories such that the equations of motion demand conformal invariance? 

This problem is technically more challenging than the open string case. In the latter, the bulk closed string CFT provided an `anchor' so that the space of boundary Lagrangians had a somewhat concrete description in terms of operators in the bulk CFT. On the other hand, the space of possible closed worldsheet theories is the space of all 2D quantum field theories! There are similar worries about the well-definedness of this space due to UV divergences and non-renormalizable field theories. 

Despite these difficulties, precisely such an action has been found by \citet{Ahmadain:2024hdp} for the classical bosonic closed string.\footnote{It is classical for similar reasons to the BSFT: It only contains the sphere partition function of the string, and so it does not account for string loops.} It has a rather unusual form: it is the product of the Zamolodchikov $C$-function (on a disk of radius $r^*$) and the partition function of the QFT on the sphere,\footnote{The Zamolodchikov $C$-function is a function which is minimised on CFTs and which returns there the central charge. For further details, see \citet{Ahmadain:2024hdp}.}
\begin{equation}\label{eq:cZ-action}
    S_{cZ} = c_{\text{pl}}(r^*)Z_{S^2}.
\end{equation}
The partition function enforces the conformal invariance condition, and the $C$-function ensures that the total central charge of the worldsheet theory is $c=0$ so that there is no conformal anomaly.\footnote{It is assumed that the ghost sector is not deformed, so the matter CFT must have $c=26$.} Both of these functions are defined for all QFTs. 

\subsubsection{Models of the $cZ$ action}\label{SS: cZ analysis}
We now analyze the models of the $cZ$ action. The space of 2D QFTs is difficult to characterize, and the interpretation of a string with an arbitrary QFT as a worldsheet theory is even more poorly understood. For this reason, we will simply restrict our attention to NLSM worldsheet theories such as those in (\ref{eqn general nl sigma model}). Just like the BSFT above, the variables of the $cZ$ action are the worldsheet couplings themselves. Thus, the models of the $cZ$ action, when restricted to this class of QFTs, can be characterized by the values of the spacetime fields which appear as couplings, $\langle M, T, g_{\mu\nu} , B_{\mu\nu}, \Phi , \dots \rangle$. These are exactly the same models that were inferred in \S \ref{SSec invariant sugra} from the background independence proofs; as mentioned there, they are the models of 26D dilaton gravity with a stringy matter sector.

\subsection{Models of string field theory}\label{SS: models of SFT}

As we have seen in this section, there are many ways in which one can characterize the models of SFT; these we have summarized in table \ref{table1}. All of these representations have benefits and drawbacks. 

The naïvely background dependent models are what are most straightforwardly read from the actions of the theories. However, as we know from formulations of gauge theories in which the symmetries are not apparent, it is not always the case that the obvious models are the physically appropriate ones. In particular, the background independence proofs from \S \ref{Sec BI} seem to suggest that certain combinations of the background fields and the string field are the true physical variables; we have called this the `inferred BI' structure. In some cases, further evidence for these models is provided by the theories in the next column, which give the same dynamics in a way that uses only these BI variables. Notably, however, such a theory is not yet available for the quantum closed string, which is ultimately the most interesting, as it is the only theory of quantum gravity in this table.


\begin{table}[] 
    \centering \footnotesize
    \begin{adjustbox}{center}
    \begin{tabular}{c|c|c|c|c}
         \textbf{Theory} & \textbf{Naïvely BD form} & \textbf{Inferred BI form} & \textbf{Manifestly BI theory} & \textbf{Manifestly BI form}  \\ \hline 
        Spin-2 gravity & $\langle M, \hat g_{\mu\nu}, h_{\mu\nu} \rangle$ & $\langle M, \hat g_{\mu\nu} +h_{\mu\nu} \rangle$ &  EH & $\langle M, g_{\mu\nu} \rangle$ \\
        Open SFT & $\langle M, \hat C, \hat O, \Psi \rangle$ & $\langle M, \hat C, \hat O + \Psi \rangle$  &  BSFT & $\langle M, \hat C, \mathcal{V} \rangle$  \\
        Classical bosonic closed SFT & $\langle M, \hat C, \Psi \rangle$ &  $\langle M, \hat C + \Psi \rangle$ &  $cZ$ (NLSM only) & $\langle M , T, g_{\mu\nu}, B_{\mu\nu}, \Phi, \dots \rangle$ \\ Quantum closed SFT  & $\langle M, \hat C, \Psi \rangle$ &  $\langle M, \hat C + \Psi \rangle$ &  ??? & ???  
    \end{tabular}
    \end{adjustbox}
    \caption{For each of (i) the spin-2 theory (our running toy example), (ii) the open SFT, (iii) the classical bosonic closed SFT, and (iv) the quantum closed SFT, (a) the naïve background dependent (`BD') structure; (b) the inferred background independent (`BI') structure; (c) the manifestly BI theory; and (d) the manifestly BI structure. By `manifestly', we here mean that dynamics (typically, an action) for the theory is provided which makes no reference to background variables or fields.}
    \label{table1}
\end{table}




\section{Assessing the background independence of string field theory}\label{Sec AssessingBI}

Having now presented and unpacked the relevant details regarding both SFT and  the core `proofs' of the background independence of SFTs, we turn now to a conceptual appraisal of the background independence of SFTs. To do so, we first review some key definitions of background independence from the philosophical literature (drawing upon a more exhaustive survey by \citet{ReadBI}) (\S\ref{SSec BIAccounts}); next, we apply these definitions to our toy case study of spin-2 gravity (\S\ref{SSec BI of spin-2}); finally, we apply the definitions to the full case of SFT, drawing upon all of the results which we have presented above (\S\ref{SSec BI of SFT}).

\subsection{Accounts of background independence}\label{SSec BIAccounts}
In assessing the background independence of SFT, we draw from the extensive literature which deals with the desideratum of background independence itself. One obvious and urgent question is this: \textit{what is} background independence? The criterion certainly captures a kind of shared intuition in the physics community, but attempts at making this intuition precise have been met with difficulties. A thorough investigation of possible answers to the question of what background independence amounts to has recently been carried out by \citet{ReadBI}, and in the following we'll draw heavily on the survey of the different accounts of background independence offered in that work.

Focusing initially on the realm of classical theories, \citet[ch.\ 3]{ReadBI} identifies a large number of different proposals for analysing the notion of background independence which have been floated in the literature. In the interests of constraining the narrative in this article while still being able to make some interesting points, we'll focus here on the following approaches to background independence:
\begin{enumerate}
    \item Absolute objects (AO)
    \item Variational principles (VP)
    \item Belot's proposal (BP)
\end{enumerate}
We now briefly introduce each of these in turn; however, we make no claim to cover these accounts in any degree of completeness. The reader interested in details of the presented accounts (such as ramifications and potential problem cases) should consult \citet[ch.\ 3]{ReadBI}.

\subsubsection{Absolute objects}
\citet[ch.\ 3]{ReadBI}, following \citet{Anderson} and later \citet{Friedman}, defines an absolute object (AO) as ``[a] geometric object which is the same (up to isomorphism) in all [dynamically possible models] of a theory.'' (It's important to be clear here that Read has in mind the `local' version of an AO proposed by \citet{Friedman}, rather than the earlier `global' version proposed by \citet{Anderson}.) The AO criterion of background independence is then captured by the following definition \citep[p.~15]{ReadBI}:
\begin{define}[\textbf{Background independence, absolute objects}]\label{Def BIAO}
    A theory is background independent iff it has no absolute objects in its formulation.
\end{define}
One of the main problems facing the AO account is that general relativity itself turns out to have an absolute object according to Anderson's definition---the metric determinant---and consequently does not qualify as background independent according to the above definition (see \citet[pp.\ 16--19]{ReadBI} for more on this point; cf.\ \citet{Pitts}).


Given these problems with the AO proposal, there are a couple of modifications which are worth mentioning. The first has to do with the notion of a `fixed field' (terminology introduced by \citet{Pooley}). According to \citet[p.~12]{ReadBI}, ``for a fixed field [...] the field values at each manifold point are identically the same in every [kinematically possible model].'' This is in contrast with absolute objects introduced above which are identical only up to isomorphism and only across dynamically possible models. The fixed field criterion of background independence is then captured by the following definition \citep[p.~20]{ReadBI}:
\begin{define}[\textbf{Background independence, fixed fields}]\label{Def BIFF}
    A theory is background independent iff it has no formulation which features fixed fields.
\end{define}
\noindent Note the crucial reliance of the fixed field criterion on the notion of theory formulation. In order for the criterion to be applicable to concrete cases, one needs to supplement it with an understanding of what it means for two theories to be equivalent. \citet[ch.\ 3]{ReadBI} considers such criteria; however, for now we only note that some such criteria are required in order to reach definite conclusions using the fixed field definition.

A third related notion is that of an `absolute field'. \citet[p.\ 22]{ReadBI} defines an absolute field as ``[a] geometric object specified in the [kinematically possible models] of a theory, which is fixed (up to isomorphism) in all [dynamically possible models] of that theory.'' This is in contrast with absolute objects (e.g.\ the metric determinant) which \textit{don't} have to be specified explicitly in the models of the theory. The absolute field criterion of background independence is then captured by the following definition \citep[p.~22]{ReadBI}:
\begin{define}[\textbf{Background independence, absolute fields}]\label{Def BIAF}
    A theory is background independent iff it has no absolute fields.
\end{define}
\noindent As discussed by \citet[ch.\ 3]{ReadBI}, the above qualification regarding the metric determinant in general relativity allows the absolute fields criterion to escape the problem case facing the absolute objects criterion; as such, it fares better than the latter in this respect, albeit at the cost of privileging specific formulations of a theory in a way which might seem \emph{ad hoc}.

\subsubsection{Variational Principles}
The account of background independence in terms of variational principles due to \citet{Pooley} differs from the ones covered so far in that it places central emphasis on the action of the theory under consideration. The proposal can be stated as follows \citep[p.~24]{ReadBI}:
\begin{define}[\textbf{Background independence, variational}]\label{Def BIVP}
    A theory is background independent iff its solution space is determined by a generally covariant action, (i) all of whose dependent variables are subject to Hamilton’s principle, and (ii) all of whose dependent variables represent physical fields.
\end{define}
Clause (ii) here plays a crucial role since, in principle, \textit{any} equation of motion can be introduced to a theory's action using Lagrange multipliers. If this proposal is to fare any better than the accounts previously canvassed, it at least needs to be able to tell apart theories which contain genuine background fields from the ones which `hide' these background fields by defining them at the level of kinematical possibilities, but subjecting them to artificial constraint equations which universally kill their variation across dynamical possibilities. Such artificial constraint equations would usually enter the action principle via unphysical Lagrange multipliers and for this reason (ii) is required. (For further discussion, see \citet[ch.\ 3]{ReadBI}.)

\subsubsection{Belot's proposal}
Belot's proposal belongs among the more involved accounts of background independence and it comes with several conceptual novelties compared to the approaches considered so far. In particular, \citet{Belot} notes that on his account:
\begin{quote}
\begin{enumerate}
    \item Background-dependence and independence come in degrees: some theories are fully background-(in)dependent, others nearly so---and others fall somewhere in between.
    \item A theory can fail to be fully background-independent in virtue of asymptotic boundary conditions.
    \item The extent of the background-(in)dependence of a theory is not a strictly formal one: in particular, it depends on how one thinks of the geometric structure of each solution and on what sorts of differences between solutions one takes to be unphysical.
\end{enumerate}
\citep[pp.\ 2872--3]{Belot}
\end{quote}
The account itself proceeds by first drawing the distinction between \textit{geometrical} and \textit{physical} degrees of freedom. In alignment with the third moral on the above list, Belot avers that whether a particular piece of structure in a physical theory qualifies as geometrical or physical is not strictly a formal matter and our interpretive convictions factor strongly into such judgements. In principle, one can take this to be an advantage of Belot's proposal; however, one should also bear in mind that definitive judgements about the background independence of a particular theory can only be drawn in conjunction with these additional interpretive judgements. Without further ado, let us now introduce the distinction between geometrical and physical degrees of freedom, following  the exposition by \citet[pp.\ 28--30]{ReadBI}.

Consider a theory $\mathcal{T}$ whose kinematically possible models are tuples of the form $\langle M, O^G, O_1,\dots,O_{n}\rangle$, where $M$ is a smooth manifold, and $O^G$ is a piece of structure identified antecedently as being `geometrical'. Such $O^G$ may in principle vary across kinematically and dynamically possible models of $\mathcal{T}$, so let us call $\mathcal{G}$ the set of all $O^G$ across all dynamically possible models of $\mathcal{T}$. Additionally, one may wish to regard certain elements of $\mathcal{G}$ as \textit{equivalent} geometrical structures and equip $\mathcal{G}$ with an equivalence relation $\sim_G$ relating equivalent geometrical structures. Taking the quotient $\mathcal{G}/\sim_G$ then defines the reduced set of geometrical objects in $\mathcal{T}$ called $\tilde{\mathcal{G}}$. Read then notes that ``[t]he degrees of freedom needed to parametrize this latter set $\tilde{\mathcal{G}}$ are what Belot calls the \textit{geometrical degrees of freedom} of $\mathcal{T}$.'' \citep[p.~29]{ReadBI}

Turning now to physical degrees of freedom, we note that it is often the case that a given physical theory commits to excess mathematical structure in the sense that two mathematically distinct models may be regarded as having exactly the same representational capacities.\footnote{There are subtleties here regarding the relations between mathematical equivalence and representational equivalence---see \citet{Fletcher2020-FLEORC}. We elide them.} Such models are called \textit{gauge-equivalent} and physical theories which contain such models are called \textit{gauge theories}.\footnote{Of course, this is just one sense in which the term `gauge theory' might be used. See \citet{Weatherall} for disambiguation.} We denote the relation of gauge equivalence by $\sim_P$, where $P$ in the subscript stands for \textit{physical} equivalence. Belot's procedure for identifying physical degrees of freedom runs as follows. Take the set $\mathcal{D}$ consisting of all the dynamically possible models of $\mathcal{T}$ and take the quotient $\tilde{\mathcal{D}} = \mathcal{D}/\sim_P$. Then ``the degrees of freedom needed to parametrize $\tilde{\mathcal{D}}$---the gauge-quotiented class of dynamically possible models is $\mathcal{T}$---are the theory's \textit{physical degrees of freedom}.'' \citep[p.~29]{ReadBI}.

Having armed ourselves with the distinction between geometrical and physical degrees of freedom, we can now formulate Belot's four definitions which characterise the background independence of a given theory:
\begin{define}[\textbf{Full background dependence, Belot}]\label{Def Belot1}
    A field theory is fully background dependent if it has no geometrical degrees of freedom: every solution is assigned the same spacetime geometry as every other solution.
\end{define}
\begin{define}[\textbf{Full background independence, Belot}]\label{Def BIBelot2}
    A field theory is fully background independent if all of its physical degrees of freedom correspond to geometrical degrees of freedom: two solutions correspond to the same physical geometry iff they are gauge equivalent.
\end{define}
\begin{define}[\textbf{Near background dependence, Belot}]\label{Def BIBelot3}
    A field theory is nearly background dependent if it has only finitely many geometrical degrees of freedom: quotienting the space of geometries that arise in solutions of the theory by the relation of geometrical equivalence yields a finite-dimensional space.
\end{define}
\begin{define}[\textbf{Near background independence, Belot}]\label{Def BIBelot4}
    A field theory is nearly background independent if it has a finite number of non-geometrical degrees of freedom: there is some $N$ such that for any geometry arising in a solution of the theory, the space of gauge equivalence classes of solutions with that geometry is no more than $N$-dimensional.
\end{define}
Again, this presentation will suffice for our purposes; see \citet[ch.\ 3]{ReadBI} for a detailed assessment and exploration of Belot's proposal. We'll see in the remainder of this section that Belot's proposal can be put to very interesting interpretive work when we asses the background independence both of the spin-2 theory and of SFT.


\subsection{Background independence of spin-2 gravity}\label{SSec BI of spin-2}

With the above definitions in hand, we'll consider now the background independence of spin-2 gravity, as a warm-up to assessing in the next subsection the background independence of SFT itself.

\subsubsection{Models of the spin-2 theory}
Having now reviewed some mainstream proposals for what background independence might consist in (as already discussed, this list is non-exhaustive; see \citet[ch.\ 3]{ReadBI} for further proposals), we can lay these proposals against our first case of interest: the spin-2 theory. Following up on \S\ref{SSec spin-2Theory}, we define the spin-2 theory as follows. Its kinematical possibilities consist of tuples $\langle M, \hat{g}_{\mu\nu}, h_{\mu\nu}\rangle$, where $M$ is a smooth manifold, $\hat{g}_{\mu\nu}$ is a Lorentzian metric and a fixed solution to the Einstein field equations on $M$, and $h_{\mu\nu}$ is a rank-2 tensor field on $M$.
The dynamically possible models are then picked out by the dynamical equations following from Hamiltonian variation with respect to $h_{\mu\nu}$ of the action functional $S_{\text{spin-2}}$ which we defined in (\ref{eqn EH-expansion}). Explicitly, from this variational procedure we obtain the equations of motion
\begin{align}\label{eqn spin-2 EoM}
    0=\frac{\delta S_{\text{spin-2}}}{\delta h_{\mu\nu}} \implies G_{\mu\nu}[\hat{g}_{\mu\nu}] + O(h_{\mu\nu}) = 0.
\end{align}
Since $\hat{g}_{\mu\nu}$ is chosen such that it solves the Einstein field equations on $M$, we see that the zeroth order term in (\ref{eqn spin-2 EoM}) vanishes. The remaining terms are then proportional to ever-increasing powers of $h_{\mu\nu}$, meaning that $h_{\mu\nu}=0$ will always be a solution of spin-2 theory, albeit a trivial one.
Formulated in this way, the spin-2 theory by construction includes the entire solution space of general relativity. Note that, on this way of thinking (which we'll adopt in the remainder of this subsection), each choice of background $\hat{g}_{\mu\nu}$ yields a \emph{distinct} spin-2 theory.

\subsubsection{Background independence of the spin-2 theory}

What should we make of the background independence of the spin-2 theory? On definition \ref{Def BIAO} (in terms of absolute objects), the theory comes out background \emph{dependent}, since it has an absolute object: the metric determinant $\det \left(  \hat{g}_{\mu\nu} \right)$ is identical across all dynamically possible models of the theory, which means that it is an absolute object in Friedman's sense. (As \citet{ReadBI} notes, this is a ubiquitous and problematic feature of the definition in terms of absolute objects: the metric determinant is an absolute object also in general relativity itself, as well as in many other spacetime theories.) Indeed, on our above way of understanding the spin-2 theory (on which each spin-2 theory has its own background $\hat{g}_{\mu\nu}$), the fixed metric $\hat{g}_{\mu\nu}$ \emph{in toto} is also an absolute object (as well as an absolute field, since $\hat{g}_{\mu\nu}$ is specified explicitly in the models of the theory---meaning that definition \ref{Def BIAF} is also violated).\footnote{If we were to have a more expansive view of the spin-2 theory according to which different, non-isomorphic backgrounds $\hat{g}_{\mu\nu}$ were permitted, then this object would not count as an absolute object---but $\det \left(  \hat{g}_{\mu\nu} \right)$ would still so count. 
Arguably, this would not be a satisfactory result, for there is a clear sense in which $\hat{g}_{\mu\nu}$ is a fixed background---namely, in the sense that it does not vary within any model, and is not coupled to any other fields. This type of case is discussed by \citet[ch.\ 3]{ReadBI} in the context of e.g.\ a cosmological constant which can vary from model to model, but which is fixed within any particular model.}\textsuperscript{,}\footnote{Are things like the Euler characteristic absolute objects? This would give the spin-2 theory an absolute object. However, this deepens the tension between including other topologies and trivializing background dependence: if there are only models of one topology, we will always have topological invariants. If not, then the fields live in completely different spaces, and it seems impossible for models of manifolds with different topologies to have an isomorphic object, thus eliminating all possibility for an absolute object.}

Let us therefore see how the remaining definitions deal with this theory. Turning to definition \ref{Def BIFF}, we note that whether or not $\hat{g}_{\mu\nu}$ is a fixed field will depend upon whether or not it is taken to be fixed \emph{identically} in all kinematical possibilities. If it is not, then this definition does not vindicate any intuitions which one might have regarding the fixed nature of $\hat{g}_{\mu\nu}$.

Definition \ref{Def BIVP} in terms of variational principles captures a clear sense in which $\hat{g}_{\mu\nu}$ is a fixed background. Recall that the spin-2 action is given by (\ref{eqn EH-expansion}) and in deriving the equations of motion (\ref{eqn spin-2 EoM}) only $h_{\mu\nu}$ is subject to Hamiltonian variation. Therefore, spin-2 theory comes out as background dependent on definition \ref{Def BIVP}.

Belot's proposal captured in definitions \ref{Def Belot1}--\ref{Def BIBelot4} deserves some unpacking in the context of the spin-2  theory. Recall that in order to put the proposal to work, one needs to identify the geometrical and physical degrees of freedom of the theory. Even at this stage, there are decisions to be made when it comes to assessing the background independence of the spin-2 theory with respect to Belot's definitions. For are the fields $\hat{g}_{\mu\nu}$ to be counted as geometrical, and are the spin-2 fields $h_{\mu\nu}$ also to be so regarded? While it seems fairly straightforward to regard $\hat{g}_{\mu\nu}$ as being `geometrical', there's surely a case to be made on either side when it comes to the spin-2 field $h_{\mu\nu}$, for on the one hand (with the origins of the spin-2 theory in general relativity in mind) the object is geometrical,  but on other it (of course) looks akin to a spin-2 material field.\footnote{It would be interesting to assess the status of $h_{\mu\nu}$ as spatiotemporal with respect to the criteria presented by \citet{MartensLehmkuhl, MartensLehmkuhl2}; for our purposes here, though, we don't need to engage in that detailed study.} And even if one does regard $h_{\mu\nu}$ as being geometrical, there are questions to be asked about whether one should be assessing the background independence of the spin-2 theory with $\hat{g}_{\mu\nu}$ and $h_{\mu\nu}$ in mind \emph{individually}, or whether should assess the background independence of the spin-2 theory with respect to the \emph{composite} object $\hat{g}_{\mu\nu} + h_{\mu\nu}$, i.e.\ the metric $g_{\mu\nu}$ of general relativity (recall \S\ref{SSec spin-2Theory}).\footnote{These issues are discussed in the context of Belot's approach to background independence by \citet[ch.\ 3]{ReadBI}.} Bringing all this together, we  have the following three cases:\footnote{In all of these cases, we're setting aside diffeomorphisms: we assume that diffeomorphisms always relate geometrically/physically equivalent models. Of course, this is somewhat controversial, in light of the points raised by \citet{BelotElvis}, and further discussed by \citet{ReadBI}. But we'll set aside the issues here.}
\begin{enumerate}
    \item \label{belot1} $\hat{g}_{\mu\nu}$ is regarded as geometrical; $h_{\mu\nu}$ is not regarded as geometrical. In this case, for each $\hat{g}_{\mu\nu}$ there are various distinct configurations of the $h_{\mu\nu}$ field, associated with the various solutions of the spin-2 theory for this background $\hat{g}_{\mu\nu}$. As such, geometrical and physical degrees of freedom do not coincide, and the theory counts as (fully) background dependent, for Belot.
    \item \label{belot2} $\hat{g}_{\mu\nu}$ is regarded as geometrical, as is $h_{\mu\nu}$, and background independence of the theory is assessed with respect to these objects, treated as fundamental. In this case, all the physical degrees of freedom are geometrical degrees of freedom, so the theory qualifies as background independent.
    \item \label{belot3} $\hat{g}_{\mu\nu}$ is regarded as geometrical, as is $h_{\mu\nu}$, and background independence of the theory is assessed with respect to the composite object $g_{\mu\nu}$, i.e.\ the metric of the associated model of general relativity. In this case, we again have it that all geometrical degrees of freedom are physical degrees of freedom, and so the theory comes out as fully background independent.
\end{enumerate}
What we find here, then, is that only on option \eqref{belot1} does the spin-2 theory come out as background dependent, on Belot's account. In the next subsection, we'll see these verdicts mirrored when we turn to the background independence of SFT.

\subsection{Background independence of SFT}\label{SSec BI of SFT}

So much for the warm-up; let's turn now to a systematic assessment of the background independence of SFTs. Our strategy for doing this will be as follows. First, in \S\ref{sec:individualtheories}, we'll assess the background independence of each individual SFT (which, as we'll see, is akin to assessing the background independence of individual spin-2 theories as discussed in the previous subsection). Then, in \S\ref{sec:invariant}, we'll assess the background independence of SFTs with respect to their invariant structure (which is akin to assessing the background independence of spin-2 theories with respect to their invariant general relativistic structure, as was also discussed in the previous subsection). And finally, in \S\ref{sec:theoriesonspace}, we'll assess the background independence of what we were calling above (prior to philosophical analysis, of course) the `manifestly background independent' formulations of these theories.

\subsubsection{Individual theories}\label{sec:individualtheories}

Our first task is to assess the background independence of an \emph{individual} SFT, with models $\langle M, \hat B, \hat \Psi\rangle$ where $\hat B$ is the fixed background and $\hat \Psi$ is the string field (recall \S\ref{SSec: Backgrounds in SFT}). As we'll see, the verdicts here mirror those which we gave for the  understanding of the spin-2 theory which we presented in the previous subsection---i.e.,\ on the understanding of spin-2 gravity according to which each background $\hat{g}_{\mu\nu}$ yields a \emph{distinct} spin-2 theory.

All of this is quite straightforward to see. Since $\hat{B}$ is fixed (up to isomorphism) across models, it at least counts as an absolute object and absolute field---although perhaps not a fixed field, since it needn't be fixed \emph{identically} in all kinematical possibilities. Still, in general one sees that an individual SFT violates `absolute object-like' definitions of background independence, as one would expect.

Consider now the definition of background independence in terms of variational principles. The first issue to settle here is the action principle which one should be considering; this is the action given in \eqref{eqn: Fock Space Expansion of string field}.
Recall that in that expression, we ignored ghost fields for simplicity, but following \citet[§10.4]{Erbin} they could be restored, and indeed for our purposes now it will be important that we consider them.

If one considers this action, then one sees that the background independence of an individual SFT appears to be violated, for our fixed, non-variational fields appear at low orders, and these fields are not subject to Hamilton's principle, from the spacetime perspective. In addition, one might think that this definition of background independence is violated doubly by virtue of the presence of the ghost fields; however, one has to be careful, because these fields \emph{are} in fact varied (see \citet[§10.4]{Erbin}). Moreover, while presumably one wouldn't want to regard these as being `physical' fields, even here there are delicate issues, as discussed by \citet{Dougherty2021-DOUIAA-2}.\footnote{It would be nice to have some principled distinction between ghosts and Lagrange multiplier fields \emph{vis-\'{a}-vis} their `physicality'. But we won't pursue this here.} And as a final point on this variational approach to background independence, note that if one considers the SFT from the \emph{worldsheet} point of view, then the fixed background fields are re-interpreted as coupling constants, and as such do not so obviously violate this definition of background independence (this point is discussed in the context of perturbative string theory by \citet[ch.\ 4]{ReadBI}). All-in-all, then, while there are interesting subtleties here regarding ghosts, what ultimately one finds is that whether an individual SFT satisfies or violates the variational definition of background independence hinges crucially on whether one adopts (respectively) the worldsheet or spacetime point of view.

Finally, we have Belot's definition of background independence. Here, \emph{prima facie} the case might seem relatively clear cut: since the background field $\hat B$---plausibly understood as the background spacetime geometry in the theory---is fixed from model to model (up to isomorphism), there are no geometrical degrees of freedom; in which case, an individual SFT is fully background dependent on Belot's account. This verdict would track case \eqref{belot1} in the previous subsection. That said, there are subtleties here again: if we take the background metric \emph{together with} the metric fluctuation of the string field as being the geometrical degrees of freedom, then in fact geometrical degrees of freedom do vary from model to model. And---assuming that these co-vary with the non-geometrical degrees of freedom (encoded in the other excitations of the string field)---one will thereby arrive at the verdict that an individual SFT is fully background \emph{independent}, in Belot's sense. This latter way of understanding the background independence of SFT in Belot's analysis is akin to case \eqref{belot2} in the previous subsection, save for the fact that we now also have material fields, encoded (as mentioned) in the other non-metrical excitations of the string field. (In addition, one might wish to regard the Kalb--Ramond field as being geometrical in string theory, since it mixes with the metric under T-duality: see \citet[\S14.2]{Blumenhagen:2013fgp}.)

\subsubsection{Invariant structure}\label{sec:invariant}

So much for assessments of the background independence of individual SFTs, interpreted `literally', where  there is a majority (but hardly unanimous) verdict that such theories are background \emph{dependent}. The next issue to consider is whether adjudications on the background independence of SFTs change when one interprets SFTs via their invariant (supergravity) structure, as discussed in \S\ref{SSec invariant sugra} and \S\ref{SS invariant YM}. Here, we'll first consider open SFT, before turning to closed SFT.

Consider first our `absolute object'-style definitions, with the invariant dynamical (super)gravity fields \eqref{eqn: sugra invariants} in mind. Here, we see (again) that definition \ref{Def BIAO} (`no absolute objects') is violated, for the usual reason that (e.g.)\ the determinant of $G_{\mu\nu}$ will count as an absolute object. On the other hand, both definition \ref{Def BIFF} and definition \ref{Def BIAF} will be satisfied, since the spacetime fields \eqref{eqn: sugra invariants} can vary (beyond isomorphism) from model to model; they are neither absolute fields nor fixed fields.

On the other hand, it is difficult to assess whether definition \ref{Def BIVP}---in terms of variational principles---is satisfied when one is focusing on these invariant structures, the reason being that (unlike in the case of the spacetime fields in perturbative string theory as considered by \citet[ch.\ 5]{ReadBI}) the action for these background (super)gravity fields is not known. Given this, it is difficult---absent further technical work---to ascertain the background independence of this `invariant' structure according to the variational principles conception of background independence. In turn, it seems that one might have to fall back on the action \eqref{eqn: Fock Space Expansion of string field} for individual SFTs, which (as we've already seen) violates definition \ref{Def BIVP}.

Finally, consider Belot's approach to background independence. With sensible choices of `geometrical' and `physical' fields (plausibly, $G_{\mu\nu}$ versus $B_{\mu\nu}$, $\Phi$, etc.---though recall our above parenthetical remark on the Kalb--Ramond field),\ one sees that geometrical degrees of freedom do indeed match physical degrees of freedom, making the theory background independent. (This mirrors the verdicts on the spacetime formulation of perturbative string theory considered by \citet[ch.\ 5]{ReadBI}).


When we turn now from the closed SFT to the open SFT, we of course still have a fixed closed string background; this background will end up violating definition \ref{Def BIAO} (absolute objects) and (plausibly) definition \ref{Def BIVP} (variational principles---to the extent that one can think about variational principles at all in the absence of a specific action). (At least on the assumption that there can't be non-isomorphic versions of this background in different models; if there in fact can be non-isomorphic versions of this background in different models, then definition \ref{Def BIAO} is in fact satisfied.) Assuming that one adopts the minor amendment to Belot's proposal for background independence made by \citet[ch.\ 3]{ReadBI}---namely, that \emph{every} object designated as `geometrical' co-vary with the `physical' degrees of freedom in one's models---these background closed string fields will also end up violating Belot's definition of background independence.

\subsubsection{Manifestly background independent formulations}\label{sec:theoriesonspace}

Finally, let's turn to what we called in table \ref{table1} the `manifestly' background independent formulations of open and closed SFTs---that is, the formulations of open and closed SFTs described by the third and fourth column of that table. (Of course, the term `manifestly' here should be taken with a pinch of salt, if one doesn't wish to cook the books in favour of a verdict of background independence---cf.\ \citet[ch.\ 5]{ReadBI}.)

Begin with Witten's BSFT for the open SFT. As before, by dint of being an open SFT, this theory has a closed SFT background in all models! (This is $\hat{C}$ in table \ref{table1}.) Typically, such a background is not understood to be fixed identically in all models, and as such it does not lead to a violation of definition \ref{Def BIAF} of background independence (in terms of fixed fields). Whether or not it violates definition \ref{Def BIAO} (absolute objects), on the other hand, will depend upon whether one countenances models of BSFT with non-isomorphic versions of $\hat{C}$---if one does this, then (mirroring our above discussion of the spin-2 theory) this field will not violate this definition of background independence; otherwise, it will lead to such a violation (in our view, mirroring the case of individual spin-2 theories, there is a case to be made that it is sensible to regard each BSFT as being associated with a specific $\hat{C}$). 
In any case, since this background is not subject to Hamilton's principle in BSFT actions such as \eqref{eq:BSFT-action}, it seems that Witten's BSFT \emph{does} violate definition \ref{Def BIVP} of background independence (variational principles). Finally, as before, whether or not Belot's definition of background independence is satisfied is again going to be a function of (a) whether different models of the BSFT can include non-isomorphic versions of $\hat{C}$, (b) whether other fields in BSFT are regarded as being `geometrical' (although the grounds on which one would regard the $\mathcal{V}$ in BSFT as being `geometrical' are not clear), and (c) whether one again adopts the above-mentioned proposed amendment to Belot's proposal made by \citet[ch.\ 3]{Read2019-REAOMA} (on which even if one were to identify other fields in BSFT as being `geometrical', the theory would still not be background independent).

Turn now to the $cZ$ action of the classical bosonic closed SFT presented by \citet{Ahmadain:2024hdp}. In this case, one has an action \eqref{eq:cZ-action} on the space of 2D CFTs which makes no reference to any fixed background whatsoever! As such, concerns regarding an outstanding fixed background which we saw for Witten's BSFT do not carry over to this case, and---\emph{prima facie} (one has to keep in mind the limitations of the approach given that the construction works for tree-level only)---one appears to have secured an approach to SFT which on almost all philosophical accounts of the notion qualifies as background independent.\footnote{One cost of the generality of the approach of \citet{Ahmadain:2024hdp} is, of course, that the physical interpretation of \eqref{eq:cZ-action} isn't obvious. But picking up this task is not our central concern here.}












\section*{Acknowledgments}

We are extremely grateful to 
Subhroneel Chakrabarti for many discussions and significant guidance over the course of this project. We're also indebted to Sebastian de Haro and to the two anonymous referees for helpful comments on a previous draft of this article. Finally, for valuable feedback we thank Nick Huggett, Chip Sebens, and audiences at Foundations of Physics 2025 (in Gda\'{n}sk) and EPSA 2025 (in Groningen).

\bibliographystyle{apalike}
 \bibliography{refs}


\end{document}